\documentclass[11pt]{article}
\usepackage{amsfonts}
\usepackage{amsgen,amsmath,amstext,amsbsy,amsopn,amssymb}
\usepackage[dvips]{graphicx}
\usepackage{comment}
\usepackage{booktabs}
\usepackage{natbib}
\usepackage[usenames]{color}
\usepackage{changepage}
\newcommand{\red}{\color{red}}

\newtheorem{theorem}{Theorem}[section]

\newtheorem{lemma}{Lemma}[section]
\newtheorem{remark}{Remark}[section]
\newtheorem{corollary}{Corollary}[section]
 
\newtheorem{proposition}{Proposition}[section]

\newcommand{\be}{\begin{equation}}
\newcommand{\ee}{\end{equation}}

 \newcommand{\beq}{\begin{equation}}
\newcommand{\eeq}{\end{equation}}
\newcommand{\beas}{\begin{eqnarray*}}
\newcommand{\eeas}{\end{eqnarray*}}
\newcommand{\bea}{\begin{eqnarray}}
\newcommand{\eea}{\end{eqnarray}}
\newcommand{\bei}{\begin{itemize}}
\newcommand{\eei}{\end{itemize}}
\newcommand{\ben}{\begin{enumerate}}
\newcommand{\een}{\end{enumerate}}
\newcommand{\bet}{\begin{theorem}}
\newcommand{\eet}{\end{theorem}}
\newcommand{\bel}{\begin{lemma}}
\newcommand{\eel}{\end{lemma}}
\newcommand{\bep}{\begin{proposition}}
\newcommand{\eep}{\end{proposition}}
\newcommand{\bed}{\begin{definition}}
\newcommand{\eed}{\end{definition}}
\newcommand{\bec}{\begin{corollary}}
\newcommand{\eec}{\end{corollary}}
\newcommand{\bex}{\begin{example}}
\newcommand{\eex}{\end{example}}

\newcommand{\B}{{\mathbf{B}}}

\newcommand{\D}{{\mathbf{D}}}
\newcommand{\I}{{\mathbf{I}}}

\newcommand{\R}{{\mathbf{R}}}

\newcommand{\Var}{{\rm Var}}

\newcommand{\X}{{\mathbf{X}}}
\newcommand{\Y}{{\mathbf{Y}}}

\newcommand{\SSigma}{\mathbf{\Sigma}}
\def\m{\boldsymbol{\mu}}

\newcommand{\T}{{\intercal}}

\newcommand{\bDelta}{\boldsymbol{\Delta}}

\newcommand{\var}{{\rm var}}

\newcommand{\diag}{{\rm diag}}

\newcommand{\argmin}{\operatornamewithlimits{argmin}}

\newcommand{\sqrtp}[1]{\left(#1\right)^{1/2}}

\begin{document}

\begin{title}
{Inference for High-dimensional Differential Correlation Matrices
\footnote{The research was supported in part by NSF Grant DMS-1208982 and NIH Grant R01 CA127334.}}
\end{title}

\author{T. Tony Cai \, and \, Anru Zhang\footnote{Corresponding author. E-mail: anrzhang@wharton.upenn.edu}\\
Department of Statistics\\
The Wharton School\\
University of Pennsylvania}
\date{} 
\maketitle

\begin{abstract}
Motivated by differential co-expression analysis in genomics, we consider in this paper estimation and testing of high-dimensional differential correlation matrices. An adaptive thresholding procedure is introduced and theoretical guarantees are given. Minimax rate of convergence is established and the proposed estimator is shown to be adaptively rate-optimal over collections of paired correlation matrices with approximately sparse differences. Simulation results show that the procedure significantly outperforms two other natural methods that are based on separate estimation of the individual correlation matrices. The procedure is also illustrated through an analysis of a breast cancer dataset, which provides evidence at the gene co-expression level that several genes, of which a subset has been previously verified, are associated with the breast cancer. 
Hypothesis testing on the differential correlation matrices is also considered. A test, which is particularly well suited for testing against sparse alternatives, is introduced.
In addition, other related problems, including estimation of a single sparse correlation matrix, estimation of the differential covariance matrices, and estimation of the differential cross-correlation matrices,  are also discussed.
\end{abstract}


\section{Introduction}\label{sec.intro}


Statistical inference on the correlation structure has a wide array of applications, ranging from  gene co-expression network analysis \citep{carter2004gene,lee2004coexpression,zhang2008class,dubois2010multiple,fuller2007weighted} to brain intelligence analysis \citep{shaw2006intellectual}.   For example, understanding the correlations between the genes  is critical for the construction of the gene co-expression network. See \cite{Kostka}, \cite{Lai}, and \cite{fuller2007weighted}. Driven by these and other applications in genomics, signal processing, empirical finance, and  many other fields,  making sound inference on the high-dimensional correlation structure  is becoming a crucial problem.

In addition to the correlation structure of a single population, the difference between the correlation matrices of two populations is of significant interest.
Differential gene expression analysis is widely used in genomics to identify disease-associated genes for complex diseases. Conventional methods mainly focus on the comparisons of the mean expression levels between the disease and control groups.  In some cases, clinical disease characteristics such as survival or tumor stage do not have significant associations with gene expression,  but there may be significant effects on gene co-expression related to the clinical outcome (\cite{shedden2005differential,hudson2009differential,bandyopadhyay2010rewiring}). Recent studies have shown that changes in the correlation networks from different stages of disease or from case and control groups are also of  importance in identifying dysfunctional gene expressions in disease.  See, for example, \cite{Fuente}.
This differential co-expression network analysis has become an important complement to the original differential expression analysis as differential correlations among the genes may reflect the rewiring of genetic networks between two different conditions  (See \cite{shedden2005differential,bandyopadhyay2010rewiring,Fuente,Ideker,fukushima2013diffcorr}). 


Motivated by these applications, we consider in this paper  optimal estimation of the differential correlation matrix.  Specifically, suppose we observe two independent sets of $p$-dimensional i.i.d. random samples $\X^{(t)} = \{\X_1^{(t)}, \ldots, \X_{n_t}^{(t)}\}$ with mean $\m_t$, covariance matrix $\SSigma_t$, and correlation matrix $\R_t$, where $t=1$ and 2.
The goal is to estimate the differential correlation matrix $\D = \R_1 - \R_2$.  A particular focus of the paper is on estimating an approximately sparse differential correlation matrix in the high dimensional setting where the dimension is much larger than the sample sizes, i.e.,  $p\gg \max(n_1, n_2)$.  The estimation accuracy is evaluated under both the spectral norm loss and the Frobenius norm loss.

A naive approach to estimating the differential correlation matrix $\D=\R_1 -\R_2$ is to first estimate the covariance matrices $\SSigma_1$ and $\SSigma_2$ separately and then normalize to obtain estimators $\hat \R_1$ and $\hat \R_2$ of  the individual correlation matrices $\R_1$ and $\R_2$, and finally take the difference
$\hat \D = \hat \R_1 - \hat \R_2$
as the estimator of the differential correlation matrix $\D$. A  simple estimate of a correlation matrix is the sample correlation matrix. However, in the high-dimensional setting, the sample correlation matrix is a poor estimate. Significant advances have been made in the last few years on optimal estimation of a high-dimensional covariance matrix.  Regularization methods such as banding, tapering, and thresholding have been proposed.  In particular, \cite{CZZ10} established the optimal rate of convergence and \cite{CY12} developed an adaptive estimator of bandable covariance matrices. For sparse covariance matrices where each row and each column has relatively few nonzero entries, \cite{Bickel_Levina} introduced a thresholding estimator and obtained rates of convergence; \cite{Cai_Liu} proposed an adaptive thresholding procedure and \cite{CZ12} established the minimax rates of convergence for estimating sparse covariance matrices. 

Structural assumptions on the individual correlation matrices $\R_1$ and $\R_2$ are crucial for the good performance of the difference estimator
. These assumptions, however, may not hold in practice. For example, gene transcriptional networks often contain the so-called hub nodes where the corresponding gene expressions are correlated with many other gene expressions. See, for example, \citep{barabasi2004network,barabasi2011network}. In such settings, some of the rows and columns of $\R_1$ and $\R_2$ have many nonzero entries which mean that $\R_1$ and $\R_2$ are not sparse. In genomic applications, the correlation matrices are rarely bandable as the genes are not ordered in any particular way.

In this paper, we propose a direct estimation method for the differential correlation matrix $\D=\R_1 - \R_2$ without first estimating $\R_1$ and $\R_2$ individually. This direct estimation method assumes that $\D$ is approximately sparse, but otherwise does not impose any structural assumptions on the individual correlation matrices  $\R_1$ and $\R_2$. An adaptive thresholding procedure is introduced and analyzed. The estimator can still perform well even when the individual correlation matrices cannot be estimated consistently. For example, direct estimation can recover the differential correlation network accurately even in the presence of hub nodes in $\R_1$ and $\R_2$ as long as the differential correlation network is approximately sparse. The key is that sparsity is assumed for $\D$ and not for $\R_1$ or $\R_2$.

Theoretical performance guarantees are provided for direct estimator of the differential correlation matrix. Minimax rates of convergence are established for the collections of paired correlation matrices with approximately sparse differences. The proposed estimator is shown to be adaptively rate-optimal. In comparison to adaptive estimation of a single sparse covariance matrix considered in  \cite{Cai_Liu}, both the procedure and the technical analysis of our method are different and more involved. Technically speaking, correlation matrix estimators are harder to analyze than those of covariance matrices and the two-sample setting in our problem further increases the difficulty.

Numerical performance of the proposed estimator is investigated through simulations. The results indicate significant advantage of estimating the differential correlation matrix directly.
The estimator outperforms two other natural alternatives that are based on separate estimation of $\R_1$ and $\R_2$.  To further illustrate the merit of the method, we apply the procedure to the analysis of a breast cancer dataset from the study by \cite{Vijver} and investigate the differential co-expressions among genes in different tumor stages of breast cancer.   
The adaptive thresholding procedure is applied to analyze the difference in the correlation alternation in different grades of tumor. 
The study provides evidence at the gene co-expression level that several genes, of which a subset has been previously verified, are associated with the breast cancer.


In addition to optimal estimation of the differential correlation matrix, we also consider  hypothesis testing of  the differential correlation matrices, $H_0: \R_1-\R_2=0$ versus $H_1: \R_1-\R_2\neq 0$. We propose a test which is particularly well suited for testing again sparse alternatives. The same ideas and techniques can also be used to treat other related problems. We also consider estimation of a single sparse correlation matrix from one random sample, estimation of the differential covariance matrices as well as estimation of the differential cross-correlation matrices.




The rest of the paper is organized as follows. Section \ref{sec.procedure} presents in detail the adaptive thresholding procedure for estimating the differential correlation matrix. The theoretical properties of the proposed estimator are analyzed in Section \ref{sec.analysis}. 
In Section \ref{sec.numeric}, simulation studies are carried out to investigate the numerical performance of the thresholding estimator and Section \ref{sec.real_data} illustrates the procedure through an analysis of a breast cancer dataset. Hypothesis testing on the differential correlation matrices is discussed in Section \ref{sec.testing}, and other related problems
are considered in the rest of Section \ref{sec.discussion}. All the proofs are given in the Appendix.

\section{Estimation of Differential Correlation Matrix}
\label{sec.procedure}

We consider in this section estimation of the differential correlation matrix and introduce a data-driven adaptive thresholding estimator. The theoretical and numerical properties of the estimator are investigated in Sections \ref{sec.analysis} and \ref{sec.numeric} respectively.

Let $\X^{(t)}=(X_1^{(t)}, \ldots, X_p^{(t)})^\T$  be a $p$-variate random vector with mean $\m_t$, covariance matrix $\SSigma_t = (\sigma_{ijt})_{1\leq i,j\leq p}$, and correlation matrix $\R_t = (r_{ijt})_{1\leq i, j\leq p}$, for $t=1$ and 2.
Suppose we observe two i.i.d. random samples, $\{\X_1^{(1)}, \ldots, \X_{n_1}^{(1)}\}$ from $\X^{(1)}$ and $\{\X_1^{(2)}, \ldots, \X_{n_2}^{(2)}\}$ from $\X^{(2)}$, and the two samples are independent. The goal is to estimate  the differential correlation matrix $\D=\R_1 - \R_2$ under the assumption that $\D$ is approximately sparse.

Given the two random samples,  the sample covariance matrices and sample correlation matrices are defined as, for $t=1$ and 2,
\bea\label{eq:hat SSigma_t}
\hat \SSigma_t &=& (\hat \sigma_{ijt})_{1\leq i, j\leq p} = \frac{1}{n_t}\sum_{k=1}^{n_t} (\X_k^{(t)} - \bar \X^{(t)})(\X_k^{(t)} - \bar \X^{(t)})^\T, \\
\label{eq:hat R_t}
\hat \R_t &=& (\hat r_{ijt})_{1\leq i, j\leq p} = \diag(\hat \SSigma_t)^{-1/2}\cdot \hat\SSigma_t \cdot \diag(\hat\SSigma_t)^{-1/2}, 
\eea
where $\bar \X^{(t)} = \frac{1}{n_t} \sum_{k=1}^{n_t} \X_k^{(t)}$ and $\diag(\hat\SSigma_t)$ is the diagonal matrix with the same diagonal as $\hat{\SSigma}_t$. 
We propose a thresholding estimator of the differential correlation matrix $\D$ by individually thresholding the entries of the difference of the two sample correlation matrices $\hat \R_1 - \hat \R_2$ with the threshold adaptive to the noise level of each entry. A key to the construction of the procedure is the estimation of the noise levels of the  individual entries of $\hat \R_1 - \hat \R_2$, as these entries are random variables themselves.

We first provide some intuition before formally introducing the estimate of the noise levels of the individual entries of $\hat \R_1 - \hat \R_2$. Note that $E\left((X_i^{(t)} - \mu_{it})(X_j^{(t)} - \mu_{jt})\right) = \sigma_{ijt}$ and $\mu_{it} \approx \bar X_i^{(t)}= \frac{1}{n_t}\sum_{k=1}^{n_t} X_{ik}$. Define
\begin{equation}\label{eq:theta_ijt}
\theta_{ijt} = \var\left((X_i^{(t)} - \mu_{it})(X_j^{(t)} - \mu_{jt})\right),\quad 1\leq i, j\leq p,\quad t = 1, 2.
\end{equation}
Then one can intuitively write 
\begin{equation}\label{eq:hat_sigma_ijt_approx}
\hat \sigma_{ijt} = \frac{1}{n_t}\sum_{k=1}^{n_t}(X_{ik}^{(t)} - \bar X_i^{(t)})(X_{jk}^{(t)} - \bar X_j^{(t)})
\approx \sigma_{ijt} + \left(\frac{\theta_{ijt}}{n_t}\right)^{1/2}z_{ijt},
\end{equation}
where $z_{ijt}$ is approximately normal with mean 0 and variance 1. Hence, $\theta_{ijt}/n_t$ measures the uncertainty of the sample covariance $\hat \sigma_{ijt}$. Based on the first order Taylor expansion of the 3-variate function $\frac{x}{(yz)^{1/2}}$ for $x\in \mathbb{R}, $ and $y, z>0$,
\begin{equation}\label{eq:taylor}
\frac{\hat x}{(\hat y \hat z)^{1/2}} = \frac{x}{(yz)^{1/2}} + \frac{\hat x - x}{(yz)^{1/2}} - \frac{x}{(yz)^{1/2}}\left(\frac{\hat y - y}{2y} + \frac{\hat z - z}{2z}\right) + o(\hat x - x) + o(\hat y - y) + o(\hat z - z),
\end{equation}
 the entries $\hat r_{ijt}$ of the sample correlation matrix $\hat\R_t = (\hat r_{ijt})$ can be approximated by
\begin{equation}\label{eq:r_ijt_approx}
\begin{split}
\hat r_{ijt} = \frac{\hat{\sigma}_{ijt}}{(\hat{\sigma}_{iit}\hat{\sigma}_{jjt})^{1/2}} & \approx  \frac{\sigma_{ijt}}{(\sigma_{iit}\sigma_{jjt})^{1/2}} + \left(\frac{\theta_{ijt}}{n_t\sigma_{iit}\sigma_{jjt}}\right)^{1/2} z_{ijt}\\
& - \frac{\sigma_{ijt}}{2(\sigma_{iit}\sigma_{jjt})^{1/2}}\left(\left(\frac{\theta_{iit}}{n_t\sigma_{iit}^2}\right)^{1/2} z_{iit} + \left(\frac{\theta_{jjt}}{n_t\sigma_{jjt}^2}\right)^{1/2}z_{jjt}\right)\\
= &\, r_{ijt} + \left(\frac{\xi_{ijt}}{n_t}\right)^{1/2} z_{ijt} - \frac{r_{ijt}}{2} \left(\left(\frac{\xi_{iit}}{n_t}\right)^{1/2}z_{iit} + \left(\frac{\xi_{jjt}}{n_t}\right)^{1/2}z_{jjt}\right),
\end{split}
\end{equation}
where we denote
\begin{equation*}
\xi_{ijt} = \frac{\theta_{ijt}}{\sigma_{iit}\sigma_{jjt}},\quad 1\leq i,j\leq p,\;  t = 1, 2.
\end{equation*}
It then follows from \eqref{eq:r_ijt_approx} that
\begin{equation}\label{eq:e_ij_approx}
\begin{split}
\hat r_{ij1} - & \hat r_{ij2} \approx  r_{ij1} - r_{ij2}  + \left(\frac{\xi_{ij1}}{n_1}\right)^{1/2} z_{ij1} - \frac{r_{ij1}}{2} \left(\left(\frac{\xi_{ii1}}{n_1}\right)^{1/2}z_{ii1} + \left(\frac{\xi_{jj1}}{n_1}\right)^{1/2}z_{jj1}\right)\\
 & - \left(\left(\frac{\xi_{ij2}}{n_2}\right)^{1/2} z_{ij2} - \frac{r_{ij2}}{2} \left(\left(\frac{\xi_{ii2}}{n_2}\right)^{1/2}z_{ii2} + \left(\frac{\xi_{jj2}}{n_2}\right)^{1/2}z_{jj2}\right)\right), \quad 1\leq i,j \leq p,
\end{split}
\end{equation}
where the random variables $z_{ij1}$ and $z_{ij2}$ are approximately normal with mean 0 and variance 1, but not necessarily independent for $1\leq i,j \leq p$. 

Equation \eqref{eq:e_ij_approx} suggests that estimation of $r_{ij1} - r_{ij2}$ is similar to the sparse covariance matrix estimation considered in \cite{Cai_Liu}, where it is proposed to adaptively threshold entries according to their individual noise levels. However, the setting here is more complicated as $\hat r_{ij1} - \hat r_{ij2}$ is not an unbiased estimate of $r_{ij1} - r_{ij2}$ and the noise levels are harder to estimate. These make the technical analysis more involved. The noise levels are unknown here but can be estimated  based on the observed data.  Specifically, we estimate $\theta_{ijt}$ and $\xi_{ijt}$ by the following data-driven quantities,
\bea\label{eq:hat_theta_ijt}
\hat\theta_{ijt} &=& {1\over n_t} \sum_{k=1}^{n_t} \left((X_{ik}^{(t)} - \bar X_i^{(t)})(X_{jk}^{(t)} - \bar X_j^{(t)}) - \hat\sigma_{ijt}\right)^2, \\
\label{eq:xi_ijt}
\hat \xi_{ijt} &=& \frac{\hat{\theta}_{ijt}}{\hat{\sigma}_{iit}\hat{\sigma}_{jjt}} = 
{1\over n_t\hat \sigma_{iit}\hat \sigma_{jjt}} \sum_{k=1}^{n_t} \left((X_{ik}^{(t)} - \bar X_i^{(t)})(X_{jk}^{(t)} - \bar X_j^{(t)}) - \hat\sigma_{ijt}\right)^2.
\eea

We are now ready to introduce the adaptive thresholding estimator of $\R_1 - \R_2$ using data-driven threshold levels. Let $s_\lambda(z)$ be a thresholding function satisfying the following conditions:
\begin{enumerate}
\item[(C1).] $|s_\lambda(z)| \leq c|y|$ for all $z, y$ satisfying $|z-y|\leq \lambda$ for some $c>0$;
\item[(C2).] $s_\lambda(z)=0$ for $|z|\leq \lambda$;
\item[(C3).] $|s_\lambda(z)-z| \leq \lambda$, for all $z\in\mathbb{R}$.
\end{enumerate}
Note that the commonly used soft thresholding function $s_\lambda(z) = \text{sgn}(z)(z-\lambda)_+$ and the adaptive lasso rule $s_\lambda = z(1 - |\lambda/z|^\eta)_+$ with $\eta \geq 1$ satisfy these three conditions. See \cite{Rothman} and \cite{Cai_Liu}. Although the hard thresholding function $s_\lambda(z) = z\cdot 1_{\{|z|\geq\lambda\}}$ does not  satisfy Condition (C1), the technical arguments given in this paper still work with very minor changes.

We propose to estimate the sparse differential correlation matrix $\D$ by the entrywise thresholding estimator $\hat \D^\ast=(\hat d_{ij}^\ast) \in\mathbb{R}^{p\times p}$  defined as
\begin{equation*}
\label{D*}
\begin{split}
\hat d^\ast_{ij} & = s_{\lambda_{ij}}(\hat r_{ij1} - \hat r_{ij2}), \quad 1\leq i,j\leq p,\\
\end{split}
\end{equation*}
where $s_\lambda(z)$ is a thresholding function satisfying (C1)-(C3) and the threshold level $\lambda_{ij}$ is given by $\lambda_{ij}= \lambda_{ij1} + \lambda_{ij2}$ with 
\begin{equation}\label{eq:lambda_ij_cor}
\lambda_{ijt} = \tau\left(\frac{\log p}{n_t}\right)^{1/2}\left(\hat\xi_{ijt}^{1/2} + \frac{|\hat r_{ijt}|}{2}\left(\hat{\xi}_{iit}^{1/2} + \hat{\xi}_{jjt}^{1/2}\right)\right),\quad 1\leq i, j\leq p, \quad t = 1, 2.
\end{equation}
Here $\hat \xi_{ijt}$ are given by \eqref{eq:xi_ijt} and the thresholding constant $\tau$ can be chosen empirically through cross-validation. See Section \ref{sec.delta} for more discussions on the empirical choice of $\tau$.

\section{Theoretical Properties}\label{sec.analysis}

We now analyze the theoretical properties of the data-driven thresholding estimator $\hat \D^\ast$ proposed in the last section. We will establish the minimax rate of convergence for estimating the differential correlation matrix $\D $ over certain classes of paired correlation matrices $(\R_1, \R_2)$ with approximately sparse difference $\D=\R_1 - \R_2$ under the spectral norm loss. The results show that $\hat \D^\ast$ is rate-optimal under mild conditions.

\subsection{Rate Optimality of the Thresholding Estimator}

We consider the following class of paired correlation matrices in $\mathbb{R}^{p\times p}$ with approximately sparse difference
\begin{equation}\label{eq:mathcal_G}
\mathcal{G}_q(s_0(p))
 =  \left\{(\R_1, \R_2): \R_1, \R_2\succeq 0; \diag(\R_1) = \diag(\R_2) = 1;  \max_i \sum_j |r_{ij1} - r_{ij2}|^q \leq s_0(p)\right\}
\end{equation}
for some $0\leq q <1$. Here $\R_1, \R_2\succeq 0$ and $\diag(\R_1) = \diag(\R_2) = 1$ mean that $\R_1$ and $\R_2$ are symmetric, semi-positive definite, and with all diagonal entries 1. For $(\R_1, \R_2)\in \mathcal{G}_q(s_0(p))$, their difference $\R_1 - \R_2$ is approximately sparse in the sense that each row vector of $\R_1 - \R_2$ lies in the $\ell_q$ ball with radius $s_0(p)$ and $0\le q<1$. 
When $q = 0$, this constraint becomes the commonly used exact sparsity condition.

Let
$$ Y_i^{(t)} = (X_i^{(t)} - \mu_{it}) / ({\rm var}(X_i^{(t)}))^{1/2}, \quad i=1,\ldots, p, \; t = 1, 2.$$
We assume that for each $i$, $Y_i$ is sub-Gaussian distributed, i.e. there exist constants $K, \eta>0$ such that for all $1\leq i \leq p$ and $t=1, 2$,
\begin{equation}\label{ineq:sub-gaussian}
Ee^{u(Y_i^{(t)})^2} \leq K , \text{ for } |u| \leq \eta.
\end{equation}
In addition, we assume for some constant $\nu_0>0$
\begin{equation}\label{ineq:var_yy}
\min_{1\leq i, j\leq p; t=1,2}\text{var} (Y_i^{(t)}Y_j^{(t)}) \geq \nu_0.
\end{equation}

The following theorem provides an upper bound for the risk of the thresholding estimator $\hat\D^\ast$ under the spectral norm loss. 

\begin{theorem}[Upper bound]\label{th:main_cor}
Suppose $\log p = o\left(\min(n_1, n_2)^{1/3}\right)$ and \eqref{ineq:sub-gaussian} and \eqref{ineq:var_yy} hold. Suppose the thresholding function $s_\lambda(z)$ satisfy Conditions (C1)-(C3). Then the thresholding estimator $\hat \D^\ast$ defined in \eqref{D*} and \eqref{eq:lambda_ij_cor} with  $\tau> 4$ satisfies
\bea\label{ineq:main_bound_spe}
\sup_{(\R_1, \R_2) \in \mathcal{G}_q(s_0(p))} E\|\hat\D^\ast - (\R_1 -\R_2)\|^2 &\leq&  
C(s_0^2(p)+1) \left(\frac{\log p}{n_1} + \frac{\log p}{n_2}\right)^{1-q}\\
\label{ineq:main_bound_l1}
\sup_{(\R_1, \R_2) \in \mathcal{G}_q(s_0(p))} E\|\hat\D^\ast - (\R_1 -\R_2)\|_{\ell_1}^2  &\leq& 
C (s_0^2(p)+1) \left(\frac{\log p}{n_1} + \frac{\log p}{n_2}\right)^{1-q}\\
\label{ineq:main_bound_F}
\sup_{(\R_1, \R_2) \in \mathcal{G}_q(s_0(p))} E\|\hat\D^\ast - (\R_1 -\R_2)\|_F^2 &\leq& 
Cp(s_0(p)+1) \left(\frac{\log p}{n_1} + \frac{\log p}{n_2}\right)^{1-q/2}
\eea
for some constant $C>0$ that does not depend on $n_1, n_2$ or $p$. 
\end{theorem}

\begin{remark} {\rm 
Condition \eqref{ineq:var_yy} holds naturally when $\X^{(t)}$ are jointly Gaussian. To see this point, we suppose $\rho_{ijt}$ is the correlation between $Y_i^{(t)}$ and $Y_j^{(t)}$. Then one can write $Y_j^{(t)} = \rho_{ijt}Y_i^{(t)} + \sqrt{1 - \rho_{ijt}^2} W$, where $Y_i^{(t)}, W$ are independently standard Gaussian. It is easy to calculate that $\Var(Y_i^{(t)}Y_j^{(t)}) = 1 + \rho_{ijt}^2 \geq 1$, which implies \eqref{ineq:var_yy} holds for $\nu_0 = 1$.
Condition \eqref{ineq:var_yy} is used in Lemma \ref{lm:Cai_Liu_lemma2} to show that  $\hat{\theta}_{ijt}$ is a good estimate of $\theta_{ijt}$ and $|\hat{\sigma}_{ijt} - \sigma_{ijt}|$ can be controlled by $C(\hat{\theta}_{ijt}\log p/n_t)^{1/2}$ with high probability. 
}
\end{remark}

Theorem \ref{th:main_cor} gives the rate of convergence for the thresholding estimator $\hat\D^\ast$. The following result provides the lower bound for the minimax risk of estimating the differential correlation matrix $\D = \R_1-\R_2$ with $(\R_1, \R_2) \in \mathcal{G}_q(s_0(p))$.
\begin{theorem}[Lower Bound]\label{pr:lower_bound} 
Suppose 
$\log p = o\left(\min (n_1, n_2)\right)$ and $s_0(p) \leq M\min(n_1, n_2)^{(1-q)/2}\\\times(\log p)^{-(3-q)/2}$
for some constant $M>0$.  Then minimax risk for estimating $\D = \R_1-\R_2$ 
satisfies
\begin{equation}\label{ineq:lower_bound_spe}
\inf_{\hat \D} \sup_{(\R_1, \R_2)\in \mathcal{G}_q(s_0(p))} E\| \hat \D - (\R_1 - \R_2)\|^2 \geq c s_0^2(p) \left(\frac{\log p}{n_1} + \frac{\log p}{n_2}\right)^{1-q},
\end{equation}
\begin{equation}\label{ineq:lower_bound_l1}
\inf_{\hat \D} \sup_{(\R_1, \R_2)\in \mathcal{G}_q(s_0(p))} E\| \hat \D - (\R_1 - \R_2)\|_{\ell_1}^2 \geq c s_0^2(p) \left(\frac{\log p}{n_1} + \frac{\log p}{n_2}\right)^{1-q},
\end{equation}
\begin{equation}\label{ineq:lower_bound_F}
\inf_{\hat \D} \sup_{(\R_1, \R_2)\in \mathcal{G}_q(s_0(p))} E\| \hat \D - (\R_1 - \R_2)\|_F^2 \geq c s_0(p) p\left(\frac{\log p}{n_1} + \frac{\log p}{n_2}\right)^{1-q/2},
\end{equation}
for some constant $c>0$.
\end{theorem}

Theorems \ref{th:main_cor} and \ref{pr:lower_bound} together yield the minimax rate of convergence 
\begin{equation*}
  s_0^2(p) \left(\frac{\log p}{n_1} + \frac{\log p}{n_2}\right)^{1-q}
\end{equation*}
for estimating $\D = \R_1-\R_2$ with $(\R_1, \R_2) \in \mathcal{G}_q(s_0(p))$ under the spectral norm loss, and show that  the thresholding estimator $\hat\D^\ast$ defined in \eqref{D*} and \eqref{eq:lambda_ij_cor} is adaptively rate-optimal.

\begin{remark}{\rm 
The technical analysis here for the different of two correlation matrices is more complicated in comparison to the problem of estimating a sparse covariance matrix considered in \cite{Cai_Liu}. It can be seen in \eqref{eq:e_ij_approx}, i.e. the ``signal + noise" expression of $\hat r_{ij1} - \hat r_{ij2}$, the difference of the sample correlation matrices has six ``noise terms". It is necessary to deal with all these six terms in the theoretical analysis of Theorem \ref{th:main_cor}.
}
\end{remark}

\section{Numerical Studies}\label{sec.numeric}

We investigate in this section the numerical performance of the adaptive thresholding estimator of the differential correlation matrix through simulations.  The method is applied to the analysis of a breast cancer dataset in the next section.

\begin{figure}[htbp]
			\centerline{\includegraphics[scale = .7]{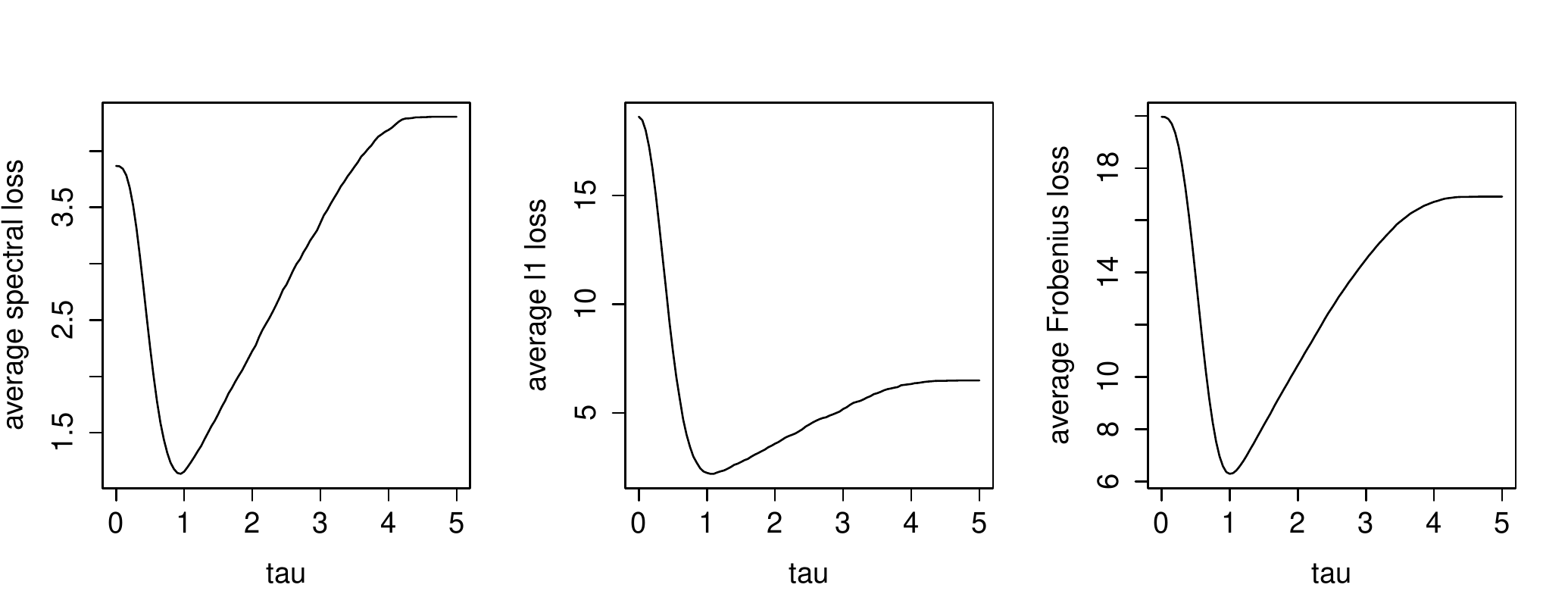}}
			\caption{Average (Spectral, $\ell_1$, Frobenious) norm losses for $\tau \in [0, 5]$. $p = 100$, $n_1 = n_2 = 50$.}
			\label{fig:tau_selection}
		\end{figure}

In the previous sections, we proposed the entrywise thresholding method for estimating $\R_1 - \R_2$ and then studied the theoretical properties of $\hat{\D}^\ast$ with  a fixed $\tau > 4$. However, the theoretical choice of $\tau$ may not be optimal in finite sample performance, as we can see in the following example. Let $\R_1$ and $\R_2$ be $200\times 200$-dimensional matrices such that $\R_{1,ij} = (-1)^{|i-j|} \times \max(1 - |i-j|/10, 0) \times (1_{\{i=j\}} + f_if_j 1_{\{i\neq j\}})$ and $\R_{2, ij} = \max(1 - |i-j|/10, 0)\times (1_{\{i=j\}} + f_if_j 1_{\{i\neq j\}})$. Here $1_{\{\cdot\}}$ is the indicator function, $f_1, \cdots, f_{200}$ are i.i.d. random variables that are uniformly distributed on $[0, 1]$. In this setting, both $\R_1$ and $\R_2$ are sparse, but their difference is even more sparse. We set $\SSigma_t = \R_t$ and generate $200$ independent samples from $\X^{(1)}\sim N(0, \SSigma_1)$ and $200$ independent samples from $\X^{(2)} \sim N(0, \SSigma_2)$. For various values of $\tau \in [0, 5]$, we implement the proposed method with hard thresholding and repeat the experiments for 100 times. The average loss in spectral, $\ell_1$ and Frobenious norms are shown in Figure \ref{fig:tau_selection}. 
Obviously in this example, $\tau>4$ is not the best choice.

Empirically, we find that the numerical performance of the estimator can often be improved by using a data-driven choice of $\tau$ based on cross-validation. We thus begin by introducing the following $K$-fold cross-validation method for the empirical selection of $\tau$.

\subsection{Empirical Choice of $\tau$}\label{sec.delta}


For an integer $K\geq 2$, we first divide both samples $\X^{(1)} = \{\X_1^{(1)}, \X_2^{(1)}, \ldots, \X_{n_1}^{(1)}\}$ and $\X^{(2)} = \{\X_1^{(2)}, \X_2^{(2)}, \ldots, \X_{n_2}^{(2)}\}$ randomly into two groups for $H$ times as $\X_{I_1^h}^{(1)}$, $\X_{T_1^h}^{(1)}$, $\X_{I_2^h}^{(2)}$ and $\X_{T_2^h}^{(2)}$. Here $h = 1, \ldots, H$ represents the $h$-th division. For $t = 1$ and $2$, the size of the first group $\X_{I_t^h}^{(t)}$ is approximately $(K-1)/K\cdot n_t$ and the size of the second group $\X_{T_t^h}^{(t)}$ is approximately $n_t/K$. We then calculate the corresponding sample correlation matrices as $\hat \R_{I^h_1}^{(1)}, \hat \R_{T^h_1}^{(1)}$, $\hat \R_{I^h_2}^{(2)}$ and $\hat \R_{T^h_2}^{(2)}$ for all four sub-samples. Partition the interval $[0, 5]$ into an equi-spaced grid $\{0, \frac{1}{N}, \ldots, \frac{5N}{N}\}$. For each value of $\tau \in \{0, \frac{1}{N}, \ldots, \frac{5N}{N}\}$, we obtain the thresholding estimator $\hat \D^\ast_{I^h}$ defined in \eqref{D*} and \eqref{eq:lambda_ij_cor} with the thresholding constant $\tau$  based on the subsamples $\X_{I_1^h}^{(1)}$ and $\X_{I_2^h}^{(2)}$. Denote the average loss for each $\tau$ for the second sub-samples $\X_{T_1^h}^{(1)}$ and $\X_{T_2^h}^{(2)}$ as
$$L(\tau) = \frac{1}{H}\sum_{h = 1}^H \|\hat\D_{I^h}^\ast - (\hat \R_{T_1^h}^{(1)} - \hat \R_{T_2^h}^{(2)})\|_F^2. $$
We select 
$$\hat \tau = \argmin_{\tau\in \left\{0, \frac{1}{N}, \ldots, \frac{5N}{N}\right\}} L(\tau)$$
as our empirical choice of the thresholding constant $\tau$, and calculate the final estimator $\hat \D^\ast(\hat \tau)$ with the thresholding constant $\hat \tau$ based on the whole samples $\X^{(1)}$ and $\X^{(2)}$.

\subsection{Estimation of Differential Correlation Matrix}

The adaptive thresholding estimator is easy to implement. We consider the following two models  under which the differential correlation matrix is sparse.

\begin{enumerate}
\item Model 1 (Random Sparse Difference)  $\R_1$ and $\R_2$ are $p$-dimensional symmetric positive definite matrices such that $\R_1 = \diag(\B_1, \I_{\frac{p}{2}\times \frac{p}{2}})$ is a fixed matrix, where $\B_1\in \mathbb{R}^{\frac{p}{2}\times \frac{p}{2}}$ with $\B_{1, ij} = 1$ if $i =j$ and $\B_{1, ij} =$0$\cdot$2 if $i\neq j$, $\I_{\frac{p}{2}\times \frac{p}{2}}$ is the $\frac{p}{2}\times \frac{p}{2}$ identity matrix, and $\R_2$ is randomly generated as $\R_2 = \diag(\B_1+\lambda \D_0,  \I_{\frac{p}{2}\times \frac{p}{2}})$, where $\D_0\in\mathbb{R}^{\frac{p}{2}\times \frac{p}{2}}$ with
$$\D_{ij, 0} = \left\{\begin{array}{cc}
1, & \text{with probability 0$\cdot$05}\\
0, & \text{with probability 0$\cdot$9}\\
-1, & \text{with probability 0$\cdot$05}
\end{array}\right. $$
and $\lambda$ is a constant that ensures  the positive definiteness of $\R_2$.

\item Model 2 (Banded Difference) In this setting, $p$-dimensional matrices $\R_1$ and $\R_2$ satisfy $\R_{1,ij} = $0$\cdot$2$\times 1_{\{i = j\}} + $0$\cdot$8$\times (-1)^{|i-j|} \times \max(1 - |i-j|/10, 0)$ and $\R_{2, ij} = \R_{1, ij} +  $0$\cdot$2$\times 1_{\{i \neq j\}}\times \max(1 - |i-j|/3, 0)$. Here $1_{\{\cdot\}}$ is the indicator function. 

\end{enumerate}
In each of the two settings,  we set $\SSigma_t = \diag(| \omega_t|^{1/2}) \R_t \diag(|\omega_t|^{1/2})$ for both $t = 1,2$, where $\omega_1, \omega_2 \in \mathbb{R}^{p}$ are two i.i.d. samples from $N(0, \I_p)$. These operations make the covariance matrices $\SSigma_1$ and $\SSigma_2$ have different values along the diagonals.


We generate i.i.d. samples from $\X^{(1)}\sim N(0, \SSigma_1)$ and $\X^{(2)}\sim N(0, \SSigma_2)$ for various values of $p, n_1$, and $n_2$ and then apply the proposed algorithm with 5-fold cross-validation for the selection of the thresholding constant $\tau$. For each setting,  both the hard thresholding and adaptive-Lasso thresholding (\cite{Rothman}),
\begin{equation}\label{eq:adaptive-Lasso}
s_\lambda(z) = z\cdot \max(1 - |\lambda/z|^\eta, 0) \quad \text{with} \quad \eta = 4,
\end{equation} 
are used.
For comparison, we also implement three natural estimators of $\D$.
\begin{enumerate}
\item The covariance matrices $\SSigma_1$ and $\SSigma_2$ are estimated individually by the adaptive thresholding method proposed in \cite{Cai_Liu} with 5-fold cross-validation and then  $\hat\SSigma^\ast_1$ and $\hat\SSigma^\ast_2$ are normalized to yield estimators of $\R_1$ and $\R_2$,
$$\hat\R_1^\ast = \diag(\hat{\SSigma}^\ast_1)^{-1/2}\hat{\SSigma}^\ast_1 \diag(\hat{\SSigma}^\ast_1)^{-1/2},\quad \hat\R_2^\ast = \diag(\hat{\SSigma}^\ast_2)^{-1/2}\hat{\SSigma}^\ast_2 \diag(\hat{\SSigma}^\ast_2)^{-1/2},$$ 
and finally  $\D=\R_1-\R_2$ is estimated by  the difference $\hat\R^\ast_1 - \hat\R^\ast_2$.

\item The correlation matrices $\hat{\R}_1^\bullet$ and $\hat{\R}_2^\bullet$ are estimated separately using the method proposed in Section \ref{sec.single_correlation} and then take the difference. 

\item $\D$ is estimated directly the difference of the sample correlation matrices $\hat \R_1 - \hat \R_2$.

\end{enumerate}

The numerical results are summarized in Tables \ref{tb:Model 1} and \ref{tb:Model 2} for the two models respectively. In each case, we compare the performance of the three estimators $\D^\ast$, $\hat\R_1^\ast - \hat\R_2^\ast$ and $\hat{\R}_1 - \hat{\R}_2$ under the spectral norm, matrix $\ell_1$ norm, and  Frobenius norm losses.
For both  models, it is easy to see that the direct thresholding estimator $\D^\ast$ significantly outperforms $\hat\R_1^\ast - \hat\R_2^\ast$ and $\hat{\R}_1 - \hat{\R}_2$. 
Under Model 1, the individual correlation matrices $\R_1$ and $\R_2$ are ``dense'' in the sense that half of the rows and columns contain many non zeros entries, but their difference $\D$ is sparse. In this case, $\R_1$ and $\R_2$ cannot be estimated consistently and the two difference estimators $\hat\R_1^\ast - \hat\R_2^\ast$ and $\hat{\R}_1 - \hat{\R}_2$ based on the individual estimators of $\R_1$ and $\R_2$ perform very poorly, while the direct estimator $\D^\ast$  performs very well. 
Moreover, the numerical performance of the thresholding estimators does not depend on the specific thresholding rules in a significant way.  Different thresholding rules including hard thresholding and adaptive Lasso behave similarly.

\begin{table}
\begin{center}
\begin{adjustwidth}{-1.5cm}{}
\begin{tabular}{cccccccccc}
\hline\hline
& & & \multicolumn{3}{c}{Hard} & \multicolumn{3}{c}{Adaptive Lasso} & Sample \\
\cmidrule(lr){4-6}\cmidrule(lr){7-9}\cmidrule(lr){10-10}
$p$ & $n_1$ & $n_2$ & $\hat{\D}^\ast$ & $\hat{\R}_1^\ast - \hat{\R}_2^\ast$ & $\hat{\R}_1^\bullet - \hat{\R}_2^\bullet$ & $\hat{\D}^\ast$ & $\hat{\R}_1^\ast - \hat{\R}_2^\ast$ & $\hat{\R}_1^\bullet - \hat{\R}_2^\bullet$ & $\hat \R_1 - \hat \R_2$ \\\hline
\multicolumn{10}{c}{Spectral Norm}\\
100 & 50 & 50 & 0$\cdot$50(0$\cdot$41) & 1$\cdot$75(1$\cdot$37) & 6$\cdot$94(1$\cdot$07) & 0$\cdot$33(0$\cdot$31) & 1$\cdot$51(1$\cdot$75) & 6$\cdot$17(0$\cdot$98) & 7$\cdot$28(0$\cdot$93)\\
100 & 100 & 100 & 0$\cdot$34(0$\cdot$21) & 3$\cdot$79(3$\cdot$17) & 4$\cdot$74(0$\cdot$55) & 0$\cdot$28(0$\cdot$23) & 3$\cdot$53(2$\cdot$71) & 4$\cdot$49(0$\cdot$71) & 5$\cdot$02(0$\cdot$65)\\
100 & 200 & 200 & 0$\cdot$29(0$\cdot$19) & 4$\cdot$22(2$\cdot$14) & 3$\cdot$23(0$\cdot$47) & 0$\cdot$24(0$\cdot$13) & 4$\cdot$52(1$\cdot$86) & 3$\cdot$14(0$\cdot$54) & 3$\cdot$55(0$\cdot$47)\\
100 & 500 & 500 & 0$\cdot$24(0$\cdot$10) & 1$\cdot$72(0$\cdot$35) & 2$\cdot$07(0$\cdot$32) & 0$\cdot$22(0$\cdot$08) & 1$\cdot$82(0$\cdot$35) & 1$\cdot$87(0$\cdot$26) & 2$\cdot$23(0$\cdot$25)\\
500 & 50 & 50 & 0$\cdot$56(0$\cdot$77) & 3$\cdot$02(2$\cdot$76) & 31$\cdot$88(4$\cdot$04) & 0$\cdot$40(0$\cdot$65) & 3$\cdot$47(4$\cdot$63) & 29$\cdot$15(4$\cdot$26) & 34$\cdot$66(3$\cdot$84)\\
500 & 100 & 100 & 0$\cdot$41(0$\cdot$48) & 8$\cdot$09(11$\cdot$02) & 23$\cdot$38(4$\cdot$52) & 0$\cdot$34(0$\cdot$39) & 12$\cdot$99(13$\cdot$26) & 21$\cdot$82(3$\cdot$23) & 24$\cdot$19(2$\cdot$77)\\
500 & 200 & 200 & 0$\cdot$32(0$\cdot$40) & 22$\cdot$22(13$\cdot$06) & 15$\cdot$67(3$\cdot$30) & 0$\cdot$26(0$\cdot$34) & 21$\cdot$31(9$\cdot$29) & 14$\cdot$61(2$\cdot$24) & 16$\cdot$50(1$\cdot$97)\\
500 & 500 & 500 & 0$\cdot$20(0$\cdot$19) & 7$\cdot$80(1$\cdot$37) & 7$\cdot$80(1$\cdot$29) & 0$\cdot$18(0$\cdot$14) & 8$\cdot$21(1$\cdot$69) & 8$\cdot$70(1$\cdot$39) & 10$\cdot$46(1$\cdot$31)\\\hline
\multicolumn{10}{c}{Matrix $\ell_1$ Norm}\\
100 & 50 & 50 & 0$\cdot$89(0$\cdot$68) & 3$\cdot$63(2$\cdot$97) & 18$\cdot$91(1$\cdot$55) & 0$\cdot$78(0$\cdot$80) & 3$\cdot$14(3$\cdot$20) & 16$\cdot$88(1$\cdot$42) & 21$\cdot$33(1$\cdot$61)\\
100 & 100 & 100 & 0$\cdot$64(0$\cdot$25) & 7$\cdot$34(4$\cdot$85) & 13$\cdot$42(1$\cdot$08) & 0$\cdot$70(0$\cdot$87) & 7$\cdot$03(4$\cdot$14) & 12$\cdot$72(1$\cdot$18) & 14$\cdot$97(1$\cdot$14)\\
100 & 200 & 200 & 0$\cdot$64(0$\cdot$34) & 9$\cdot$63(1$\cdot$85) & 9$\cdot$22(0$\cdot$75) & 0$\cdot$61(0$\cdot$37) & 9$\cdot$37(1$\cdot$77) & 8$\cdot$67(0$\cdot$87) & 10$\cdot$60(0$\cdot$81)\\
100 & 500 & 500 & 0$\cdot$58(0$\cdot$22) & 4$\cdot$54(0$\cdot$56) & 5$\cdot$92(0$\cdot$54) & 0$\cdot$56(0$\cdot$21) & 4$\cdot$85(0$\cdot$61) & 5$\cdot$33(0$\cdot$45) & 6$\cdot$69(0$\cdot$44)\\
500 & 50 & 50 & 1$\cdot$69(3$\cdot$09) & 7$\cdot$85(8$\cdot$14) & 97$\cdot$28(6$\cdot$64) & 1$\cdot$37(2$\cdot$87) & 9$\cdot$49(11$\cdot$93) & 87$\cdot$02(7$\cdot$31) & 112$\cdot$40(8$\cdot$97)\\
500 & 100 & 100 & 1$\cdot$06(1$\cdot$28) & 20$\cdot$12(19$\cdot$50) & 64$\cdot$98(5$\cdot$93) & 1$\cdot$17(1$\cdot$47) & 27$\cdot$95(22$\cdot$75) & 65$\cdot$60(5$\cdot$07) & 79$\cdot$66(5$\cdot$37)\\
500 & 200 & 200 & 0$\cdot$97(1$\cdot$48) & 51$\cdot$64(10$\cdot$13) & 46$\cdot$74(5$\cdot$01) & 0$\cdot$95(1$\cdot$24) & 47$\cdot$87(8$\cdot$50) & 45$\cdot$54(3$\cdot$70) & 55$\cdot$77(3$\cdot$70)\\
500 & 500 & 500 & 0$\cdot$68(0$\cdot$54) & 23$\cdot$19(2$\cdot$86) & 23$\cdot$47(2$\cdot$56) & 0$\cdot$69(0$\cdot$52) & 24$\cdot$67(2$\cdot$92) & 27$\cdot$21(2$\cdot$09) & 35$\cdot$32(2$\cdot$04)\\
\hline
\multicolumn{10}{c}{Frobenious Norm}\\
100 & 50 & 50 & 1$\cdot$40(1$\cdot$32) & 4$\cdot$34(2$\cdot$51) & 19$\cdot$01(0$\cdot$37) & 1$\cdot$06(1$\cdot$04) & 3$\cdot$26(2$\cdot$61) & 16$\cdot$60(0$\cdot$42) & 19$\cdot$87(0$\cdot$38)\\
100 & 100 & 100 & 0$\cdot$96(0$\cdot$59) & 7$\cdot$14(3$\cdot$67) & 13$\cdot$38(0$\cdot$23) & 0$\cdot$94(0$\cdot$81) & 6$\cdot$23(3$\cdot$31) & 12$\cdot$10(0$\cdot$27) & 14$\cdot$05(0$\cdot$25)\\
100 & 200 & 200 & 0$\cdot$89(0$\cdot$57) & 9$\cdot$09(1$\cdot$03) & 9$\cdot$30(0$\cdot$20) & 0$\cdot$84(0$\cdot$50) & 8$\cdot$15(1$\cdot$07) & 8$\cdot$42(0$\cdot$25) & 9$\cdot$94(0$\cdot$18)\\
100 & 500 & 500 & 0$\cdot$85(0$\cdot$32) & 4$\cdot$37(0$\cdot$27) & 5$\cdot$92(0$\cdot$11) & 0$\cdot$82(0$\cdot$30) & 4$\cdot$43(0$\cdot$25) & 5$\cdot$26(0$\cdot$11) & 6$\cdot$39(0$\cdot$10)\\
500 & 50 & 50 & 3$\cdot$33(5$\cdot$63) & 11$\cdot$18(7$\cdot$71) & 95$\cdot$05(0$\cdot$91) & 2$\cdot$27(3$\cdot$92) & 9$\cdot$57(9$\cdot$31) & 83$\cdot$40(1$\cdot$24) & 99$\cdot$97(0$\cdot$93)\\
500 & 100 & 100 & 2$\cdot$18(2$\cdot$98) & 20$\cdot$17(14$\cdot$55) & 61$\cdot$37(2$\cdot$09) & 2$\cdot$10(2$\cdot$47) & 22$\cdot$54(16$\cdot$69) & 60$\cdot$58(0$\cdot$85) & 70$\cdot$40(0$\cdot$67)\\
500 & 200 & 200 & 1$\cdot$77(2$\cdot$39) & 45$\cdot$06(5$\cdot$89) & 42$\cdot$11(1$\cdot$67) & 1$\cdot$63(1$\cdot$96) & 39$\cdot$52(5$\cdot$14) & 41$\cdot$88(0$\cdot$73) & 49$\cdot$53(0$\cdot$49)\\
500 & 500 & 500 & 1$\cdot$27(1$\cdot$09) & 20$\cdot$17(0$\cdot$99) & 21$\cdot$74(0$\cdot$65) & 1$\cdot$22(0$\cdot$84) & 20$\cdot$48(0$\cdot$91) & 25$\cdot$47(0$\cdot$41) & 31$\cdot$34(0$\cdot$33)\\\hline
\end{tabular}
\end{adjustwidth}
\end{center}
\caption{Comparison of $\hat \D^\ast$ with $\hat \R_1^\ast - \hat{\R}_2^\ast$ and $\hat \R_1 - \hat{\R}_2$ under Model 1.}
\label{tb:Model 1}
\end{table}

\begin{table}
\begin{center}
\begin{adjustwidth}{-1.5cm}{}
\begin{tabular}{cccccccccc}
\hline\hline
& & & \multicolumn{3}{c}{Hard} & \multicolumn{3}{c}{Adaptive Lasso} & Sample \\
\cmidrule(lr){4-6}\cmidrule(lr){7-9}\cmidrule(lr){10-10}
$p$ & $n_1$ & $n_2$ & $\hat{\D}^\ast$ & $\hat{\R}_1^\ast - \hat{\R}_2^\ast$ & $\hat{\R}_1^\bullet - \hat{\R}_2^\bullet$ & $\hat{\D}^\ast$ & $\hat{\R}_1^\ast - \hat{\R}_2^\ast$ & $\hat{\R}_1^\bullet - \hat{\R}_2^\bullet$ & $\hat \R_1 - \hat \R_2$ \\\hline
\multicolumn{10}{c}{Spectral Norm}\\
100 & 50 & 50 &  0$\cdot$98(1$\cdot$00) & 4$\cdot$61(1$\cdot$49) & 7$\cdot$25(0$\cdot$87) & 0$\cdot$71(0$\cdot$70) & 4$\cdot$47(1$\cdot$44) & 6$\cdot$05(0$\cdot$74) & 8$\cdot$29(0$\cdot$98)\\
100 & 100 & 100 & 0$\cdot$70(0$\cdot$51) & 2$\cdot$88(0$\cdot$81) & 5$\cdot$01(0$\cdot$57) & 0$\cdot$62(0$\cdot$47) & 2$\cdot$93(0$\cdot$87) & 4$\cdot$25(0$\cdot$52) & 5$\cdot$83(0$\cdot$59) \\
100 & 200 & 200 & 0$\cdot$60(0$\cdot$35) & 1$\cdot$93(0$\cdot$55) & 3$\cdot$53(0$\cdot$42) & 0$\cdot$48(0$\cdot$24) & 1$\cdot$98(0$\cdot$57) & 2$\cdot$98(0$\cdot$37) & 4$\cdot$07(0$\cdot$47) \\
100 & 500 & 500 & 0$\cdot$47(0$\cdot$14) & 1$\cdot$23(0$\cdot$27) & 2$\cdot$32(0$\cdot$23) & 0$\cdot$46(0$\cdot$17) & 1$\cdot$30(0$\cdot$36) & 1$\cdot$98(0$\cdot$21) & 2$\cdot$66(0$\cdot$27)\\
500 & 50 & 50 & 0$\cdot$97(0$\cdot$99) & 5$\cdot$03(1$\cdot$00) & 20$\cdot$61(1$\cdot$07) & 0$\cdot$80(0$\cdot$75) & 4$\cdot$55(0$\cdot$96) & 16$\cdot$91(0$\cdot$90) & 24$\cdot$96(1$\cdot$16)\\
500 & 100 & 100 & 0$\cdot$79(0$\cdot$62) & 3$\cdot$17(0$\cdot$49) & 13$\cdot$64(0$\cdot$59) & 0$\cdot$59(0$\cdot$41) & 3$\cdot$14(0$\cdot$63) & 11$\cdot$13(0$\cdot$53) & 16$\cdot$39(0$\cdot$71)\\
500 & 200 & 200 & 0$\cdot$60(0$\cdot$36) & 2$\cdot$13(0$\cdot$30) & 9$\cdot$12(0$\cdot$42) & 0$\cdot$51(0$\cdot$31) & 2$\cdot$11(0$\cdot$35) & 7$\cdot$44(0$\cdot$37) & 10$\cdot$94(0$\cdot$50) \\
500 & 500 & 500 & 0$\cdot$51(0$\cdot$20) & 1$\cdot$34(0$\cdot$16) & 5$\cdot$63(0$\cdot$23) & 0$\cdot$49(0$\cdot$20) & 1$\cdot$35(0$\cdot$22) & 4$\cdot$65(0$\cdot$21) & 6$\cdot$78(0$\cdot$29)\\\hline
\multicolumn{10}{c}{Matrix $\ell_1$ Norm}\\
100 & 50 & 50 & 1$\cdot$84(2$\cdot$66) & 10$\cdot$61(3$\cdot$48) & 19$\cdot$11(1$\cdot$55) & 1$\cdot$26(1$\cdot$86) & 9$\cdot$88(3$\cdot$14) & 16$\cdot$18(1$\cdot$55) & 21$\cdot$92(1$\cdot$61)\\
 100 & 100 & 100 & 1$\cdot$18(1$\cdot$44) & 6$\cdot$73(2$\cdot$24) & 13$\cdot$62(1$\cdot$08) & 1$\cdot$10(1$\cdot$26) & 6$\cdot$73(2$\cdot$12) & 11$\cdot$73(1$\cdot$20) & 15$\cdot$87(1$\cdot$11) \\
100 & 200 & 200 & 0$\cdot$98(0$\cdot$98) & 4$\cdot$53(1$\cdot$47) & 9$\cdot$79(0$\cdot$85) & 0$\cdot$71(0$\cdot$71) & 4$\cdot$68(1$\cdot$46) & 8$\cdot$39(0$\cdot$89) & 11$\cdot$37(0$\cdot$97)\\
100 & 500 & 500 & 0$\cdot$67(0$\cdot$48) & 2$\cdot$95(0$\cdot$89) & 6$\cdot$44(0$\cdot$56) & 0$\cdot$65(0$\cdot$59) & 3$\cdot$08(1$\cdot$03) & 5$\cdot$58(0$\cdot$47) & 7$\cdot$47(0$\cdot$53)\\
500 & 50 & 50 & 1$\cdot$79(2$\cdot$65) & 11$\cdot$03(2$\cdot$80) & 79$\cdot$71(2$\cdot$64) & 1$\cdot$64(2$\cdot$26) & 10$\cdot$38(3$\cdot$02) & 64$\cdot$46(2$\cdot$73) & 97$\cdot$88(2$\cdot$55)\\
500 & 100 & 100 & 1$\cdot$45(1$\cdot$75) & 7$\cdot$66(1$\cdot$79) & 56$\cdot$52(2$\cdot$16) & 1$\cdot$02(1$\cdot$35) & 7$\cdot$73(2$\cdot$40) & 45$\cdot$65(1$\cdot$86) & 69$\cdot$42(1$\cdot$76)\\
500 & 200 & 200 & 1$\cdot$02(1$\cdot$18) & 4$\cdot$97(1$\cdot$09) & 39$\cdot$86(1$\cdot$39) & 0$\cdot$83(1$\cdot$14) & 5$\cdot$03(1$\cdot$27) & 31$\cdot$90(1$\cdot$15) & 49$\cdot$11(1$\cdot$33) \\
500 & 500 & 500 & 0$\cdot$81(0$\cdot$70) & 3$\cdot$15(0$\cdot$72) & 25$\cdot$34(0$\cdot$77) & 0$\cdot$82(0$\cdot$77) & 3$\cdot$27(1$\cdot$00) & 20$\cdot$39(0$\cdot$77) & 31$\cdot$36(0$\cdot$79)\\\hline
\multicolumn{10}{c}{Frobenious Norm}\\
100 & 50 & 50 & 3$\cdot$36(2$\cdot$53) & 13$\cdot$82(1$\cdot$83) & 18$\cdot$46(0$\cdot$81) & 2$\cdot$66(1$\cdot$47) & 12$\cdot$13(2$\cdot$00) & 15$\cdot$87(0$\cdot$79) & 19$\cdot$92(0$\cdot$94)\\
100 & 100 & 100 & 2$\cdot$67(1$\cdot$19) & 9$\cdot$46(1$\cdot$28) & 13$\cdot$26(0$\cdot$55) & 2$\cdot$54(1$\cdot$10) & 8$\cdot$77(1$\cdot$18) & 11$\cdot$51(0$\cdot$54) & 14$\cdot$32(0$\cdot$58) \\
100 & 200 & 200 & 2$\cdot$43(0$\cdot$69) & 6$\cdot$94(0$\cdot$72) & 9$\cdot$75(0$\cdot$39) & 2$\cdot$26(0$\cdot$51) & 6$\cdot$68(0$\cdot$76) & 8$\cdot$59(0$\cdot$37) & 10$\cdot$49(0$\cdot$43)\\
100 & 500 & 500 & 2$\cdot$24(0$\cdot$34) & 5$\cdot$29(0$\cdot$36) & 6$\cdot$96(0$\cdot$19) & 2$\cdot$25(0$\cdot$46) & 5$\cdot$28(0$\cdot$44) & 6$\cdot$33(0$\cdot$17) & 7$\cdot$39(0$\cdot$19)\\
500 & 50 & 50 & 6$\cdot$77(4$\cdot$86) & 34$\cdot$24(3$\cdot$33) & 91$\cdot$09(0$\cdot$85) & 6$\cdot$18(3$\cdot$86) & 27$\cdot$39(3$\cdot$39) & 75$\cdot$97(0$\cdot$83) & 100$\cdot$71(0$\cdot$92)\\
500 & 100 & 500 & 6$\cdot$19(2$\cdot$98) & 22$\cdot$76(1$\cdot$92) & 64$\cdot$37(0$\cdot$56) & 5$\cdot$30(1$\cdot$79) & 20$\cdot$12(2$\cdot$21) & 53$\cdot$72(0$\cdot$56) & 71$\cdot$23(0$\cdot$58)\\
500 & 200 & 200 & 5$\cdot$32(1$\cdot$49) & 16$\cdot$34(1$\cdot$18) & 45$\cdot$79(0$\cdot$44) & 5$\cdot$10(1$\cdot$36) & 15$\cdot$01(1$\cdot$15) & 38$\cdot$36(0$\cdot$42) & 50$\cdot$61(0$\cdot$43) \\
500 & 500 & 500 & 5$\cdot$00(0$\cdot$62) & 12$\cdot$14(0$\cdot$59) & 29$\cdot$77(0$\cdot$27) & 4$\cdot$99(0$\cdot$69) & 11$\cdot$76(0$\cdot$70) & 25$\cdot$27(0$\cdot$24) & 32$\cdot$80(0$\cdot$25) \\\hline
\end{tabular}
\end{adjustwidth}
\end{center}
\label{tb:Model 2}
\caption{Comparison of $\hat \D^\ast$ with $\hat \R_1^\ast - \hat{\R}_2^\ast$ and $\hat \R_1 - \hat{\R}_2$ under Model 2.}
\end{table}

\section{Analysis of A Breast Cancer Dataset}
\label{sec.real_data}


Identifying gene expression networks can be helpful for conducting more effective treatment based to the condition of patients. \cite{Fuente} demonstrated that the gene expression networks can vary in different disease states and the differential correlations in gene expression (i.e. co-expression) are useful in disease studies. 

In this section, we consider the dataset ``70pathwaygenes-by-grade" from the study by \cite{Vijver} and investigate the differential co-expressions among genes in different tumor stages of breast cancer. In this dataset, there are 295 records of patients with 1624 gene expressions, which are categorized into three groups based on the histological grades of tumor (``Good", ``Intermediate" and ``Poor") with 74, 101 and 119 records, respectively. We denote these three groups of samples as $\X^{(1)}, \X^{(2)}$ and $\X^{(3)}$. In order to analyze the difference in the correlation alternation in different grades of tumor, we apply our adaptive thresholding method with cross-validation to estimate the differential correlation matrices among those gene expressions from different stages. 

The number of gene pairs with significant difference in correlation are listed in Table \ref{tb:number of selected pairs}. The results show that the correlation structures between the ``Good" and ``Intermediate" groups are similar and there is some significant changes between the  ``Good" and ``Poor" group.

\begin{table}
\begin{center}
\begin{tabular}{cccc}
\hline
& Good v.s. Intermediate & Intermediate v.s. Poor & Good v.s. Poor \\ \hline
\# of selected pairs & 0 & 2 & 152 \\ \hline
\end{tabular}
\end{center}
\caption{The number of gene pairs that have significant differential correlation betweens two groups of different tumor grades}
\label{tb:number of selected pairs}
\end{table}

More interestingly, by combining the ``Good" and ``Intermediate" sub-samples and comparing with the ``Poor" group, we find significant differences between their correlation structure. There are 4526 pairs of genes that have significantly different correlations  between the ``Good + Intermediate" and ``Poor" groups.  For each given gene, we count the number of the genes whose correlation with this gene is significantly different between  these two groups, and rank all the genes by the counts. That is, we rank the genes by the size of the support of $\hat \D^\ast$ in each row.
The top ten genes are listed in Table \ref{tb:genes}. 

\begin{table}[h]
\begin{center}
\begin{tabular}{lc}
\hline
Gene & number of pairs\\\hline
growth differentiation factor 5 (GDF5) & 67 \\
transcription factor 7-like 1 (TCF7L1) & 64 \\
3'-phosphoadenosine 5'-phosphosulfate synthase 1 (PAPSS1) & 51 \\
secreted frizzled-related protein 1(SFRP1) & 43 \\
 gamma-aminobutyric acid A receptor, pi (GABRP) & 41 \\ 
mannosidase, alpha, class 2B, member 2 (MAN2B2) & 37 \\
 desmocollin 2 (DSC2) & 36 \\
 transforming growth factor, beta 3 (TGFB3) & 35 \\
  CRADD & 35 \\
 ELOVL fatty acid elongase 5(ELOVL5) & 32 \\\hline
\end{tabular}
\end{center}
\caption{The top ten genes that appear for most times in the selected pairs in ``Good + Intermediate" v.s. ``Poor"}
\label{tb:genes}
\end{table}

Among these ten genes, six of them, GDF5, TCF7L1, PAPSS1, SFRP1, GABRP, TGFB1, have been previously studied and verified in the literature that are associated with the breast cancer (See \cite{GDF5}, \cite{Tcf7l1}, \cite{PAPSS1}, \cite{SFRP1}, \cite{GABRP},  and \cite{TGFB1}, respectively). Take for example GDF5 and TCF7L1, the overproduction of Transforming growth factor beta-1 (TGF$\beta$), a multifunctional cytokine, is an important characteristic of late tumor progression. Based on the study by \cite{GDF5}, TGF{$\beta$}  produced by breast cancer cells brings about in endothelial cells expression of GDF5. The findings in  (\cite{Tcf7l1}) suggested the important role played by TCF7L1 in breast cancer. Although these biological studies mainly focus on the the behavior of the single gene expression, our study provides evidence in the gene co-expression level that these gene expressions are related with the breast cancer.

We should point out that the two well-known genes related to the breast cancer, BRCA1 and BRCA2, were not detected by our method. This is mainly due to the fact that our method focus on the differential gene co-expressions, not the changes in the gene expression levels.

\section{Other Related Problems}
\label{sec.discussion}

We have so far focused on optimal estimation of the differential correlation matrix. In addition to optimal estimation, hypothesis testing of the differential correlation matrix is also an important problem. In this section we consider testing the hypotheses $H_0: \R_1-\R_2=0$ versus $H_1: \R_1-\R_2\neq 0$ and propose a test which is particularly well suited for testing again sparse alternatives. 

Similar ideas and techniques can also be used to treat several other related problems, including estimation of a single sparse correlation matrix from one random sample, estimation of the differential covariance matrices, and estimation of the differential cross-correlation matrices. We also briefly discuss these problems in this section.

\subsection{Testing Differential Correlation Matrices}
\label{sec.testing}

Suppose we are given two sets of independent and identical distributed samples $\X^{(t)} = \{\X_1^{(t)}, \ldots, \X_{n_t}^{(t)}\}$ with the mean $\mu_t$, covariance matrix $\SSigma_t$ and correlation matrix $\R_t$, where $t = 1$ and $2$, and wish to test the hypotheses
\begin{equation}\label{eq:test}
H_0:\R_1 - \R_2=0 \quad \text{v.s.} \quad H_1: \R_1 -\R_2 \neq 0.
\end{equation}
This testing problem is similar to, but also different from, testing the equality of two high-dimensional covariance matrices, which has been considered in several recent papers. See, for example, \cite{schott2007test}, \cite{srivastava2010testing}, \cite{li2012two}, and \cite{Cai_Liu_Xia}.
Here we are particularly interested in testing against sparse alternatives and follow similar ideas as those in \cite{Cai_Liu_Xia}.

To construct the test statistic, we need more precise understanding of  the sample correlation coefficients $\hat r_{ijt}$. It follows from \eqref{eq:taylor} that
\begin{equation*}
\begin{split}
\hat r_{ijt}  = & \frac{\hat{\sigma}_{ijt}}{(\hat{\sigma}_{iit}\hat{\sigma}_{jjt})^{1/2}}
 \approx  \frac{\sigma_{ijt}}{(\sigma_{iit}\sigma_{jjt})^{1/2}} + \frac{\hat{\sigma}_{ijt} - \sigma_{ijt}}{(\sigma_{iit}\sigma_{jjt})^{1/2}} - \frac{\sigma_{ijt}}{2(\sigma_{iit}\sigma_{jjt})^{1/2}}\left(\frac{\hat{\sigma}_{iit} - \sigma_{iit}}{(\sigma_{iit}\sigma_{iit})^{1/2}} + \frac{\hat{\sigma}_{jjt} - \sigma_{jjt}}{(\sigma_{jjt}\sigma_{jjt})^{1/2}}\right)\\
 = & r_{ijt} + \frac{1}{n_t}\sum_{k=1}^{n_t}\Bigg[ \frac{(X_{ik}^{(t)} - \bar X_i^{(t)})(X_{jk}^{(t)} - \bar X_j^{(t)}) - \sigma_{ijt}}{(\sigma_{iit}\sigma_{jjt})^{1/2}}\\ 
 & \hskip1in~ - \frac{r_{ijt}}{2}\left(\frac{(X_{ik}^{(t)} - \bar X_i^{(t)})^2 - \sigma_{iit}}{\sigma_{iit}} + \frac{(X_{jk}^{(t)} - \bar X_j^{(t)})^2 - \sigma_{jjt}}{\sigma_{jjt}}\right)\Bigg]
\end{split}
\end{equation*}
Since $\bar X_i^{(t)} \approx \mu_{it}$, $\bar X_j^{(t)} \approx \mu_{jt}$, $E\left(X_{ik}^{(t)} - \bar X_{i}^{(t)}\right)\left(X_{jk}^{(t)} - \bar X_j^{(t)}\right)  \approx \sigma_{ijt}$, we introduce
\begin{equation*}
\eta_{ijt} = \var\left[\frac{(X_i^{(t)} - \mu_{it})(X_j^{(t)} - \mu_{jt})}{(\sigma_{iit}\sigma_{jjt})^{1/2}} - \frac{r_{ijt}}{2}\left(\frac{(X_i^{(t)} - \mu_{it})^2}{\sigma_{iit}} + \frac{(X_j^{(t)} - \mu_{jt})^2}{\sigma_{jjt}}\right)\right].
\end{equation*}
Then asymptotically as $n, p\to \infty$,
\begin{equation*}
\hat r_{ijt} - r_{ijt} \approx \left(\frac{\eta_{ijt}}{n_t}\right)^{1/2} z_{ijt} ,\quad\mbox{where}\quad z_{ijt} \sim N(0, 1).
\end{equation*}
The true value of $\eta_{ijt}$ is unknown but can be estimated by
\begin{equation*}
\begin{split}
\hat{\eta}_{ijt} = & \frac{1}{n_t}\sum_{k=1}^{n_t} \Bigg\{\frac{(X_{ik}^{(t)} - \bar X_i^{(t)})(X_{jk} - \bar X_j^{(t)}) - \hat \sigma_{ijt}}{(\hat\sigma_{iit}\hat\sigma_{jjt})^{1/2}}\\
& - \frac{\hat r_{ijt}}{2}\left(\frac{(X_{ik}^{(t)} - \bar X_i^{(t)})^2 - \hat{\sigma}_{iit}}{\hat\sigma_{iit}} + \frac{(X_{jk}^{(t)} - \bar X_j^{(t)})^2 - \hat{\sigma}_{jjt}}{\hat\sigma_{jjt}}\right)\Bigg\}^2\\
= & \frac{1}{n_t}\sum_{k=1}^{n_t} \left\{\frac{(X_{ik}^{(t)} - \bar X_i^{(t)})(X_{jk} - \bar X_j^{(t)})}{(\hat\sigma_{iit}\hat\sigma_{jjt})^{1/2}} - \frac{\hat r_{ijt}}{2}\left(\frac{(X_{ik}^{(t)} - \bar X_i^{(t)})^2}{\hat\sigma_{iit}} + \frac{(X_{jk}^{(t)} - \bar X_j^{(t)})^2}{\hat\sigma_{jjt}}\right)\right\}^2.
\end{split}
\end{equation*}
We define the test statistic by
\begin{equation*}
T_n = \max_{1\leq i\leq j\leq p} T_{ij} 
\end{equation*}
where
\begin{equation*}
T_{ij} = \frac{(\hat r_{ij1} - \hat r_{ij2})^2}{\hat{\eta}_{ij1}/n_1 + \hat{\eta}_{ij2}/n_2}, \quad 1\leq i, j\leq p.
\end{equation*}
Under regularity conditions (similar to (C1)-(C3) in \cite{Cai_Liu_Xia}), the asymptotic distribution of $T_n$ can be shown to be  the type I extreme value distribution. More precisely,
\begin{equation}\label{eq:null_distribution}
P\left(T_n - 4\log p + \log \log p \leq  t\right) \to \exp\left(-(8\pi)^{-1/2}\exp\left(-t/2\right)\right)
\end{equation}
for any given $t\in \mathbb{R}$.  

The asymptotic null distribution \eqref{eq:null_distribution} can then be used to construct a test for testing the hypothesis $H_0:\R_1 - \R_2 =0$. 
For a given significance level $0<\alpha < 1$, define the test $\Psi_\alpha$ by
\beq
\label{test}
\Psi_{\alpha}= I( M_{n}\geq 4\log p-\log\log p + \tau_{\alpha})
\eeq
where $\tau_\alpha = -\log (8\pi) - 2 \log\log(1-\alpha)^{-1}$
is the $1-\alpha$ quantile of  the type I extreme value distribution with the cumulative distribution function $\exp(-(8\pi)^{-1/2}\exp(-x/2))$. The hypothesis $H_0: \R_{1}-\R_{2}=0$ is rejected whenever $\Psi_\alpha =1$.
As the test proposed in \cite{Cai_Liu_Xia} for testing the equality of two covariance matrices, the test $\Psi_\alpha$ defined in \eqref{test} can also be shown to be particularly well suited for testing $H_0:\R_1 - \R_2=0 $ against sparse alternatives.

\subsection{Optimal Estimation of a Sparse Correlation Matrix}\label{sec.single_correlation}

The ideas and technical tools can also be used for estimation of a single correlation matrix from one random sample, which is a simpler problem. 
Suppose we observe an independent and identical distributed sample $\X = (\X_1, \ldots, \X_n)$ from a $p$-dimensional distribution with mean $\m \in\mathbb{R}^{p}$, covariance matrix $\SSigma$,  and correlation matrix $\R\in\mathbb{R}^{p\times p}$.
When $\R$ is approximately sparse, it can be naturally estimated by a thresholding estimator $\hat \R$ as follows. 
Let $\bar X_i = {1\over n} \sum_{k=1}^nX_{ik}$. Define the sample covariance matrix $\hat{\SSigma}= (\hat \sigma_{ij})_{1\leq i, j\leq p}$ and the sample correlation matrix  $\hat{\R} = (\hat r_{ij})_{1\leq i,j\leq p}$ respectively by
\begin{equation*}
\hat{\sigma}_{ij} = \frac{1}{n} \sum_{k=1}^{n}(X_{ik} - \bar X_i)(X_{jk} - \bar X_j)
\quad\mbox{and}\quad 
\hat r_{ij} = \frac{\hat{\sigma}_{ij}}{(\hat{\sigma}_{ii}\hat{\sigma}_{jj})^{1/2}}.
\end{equation*}
Same as in \eqref{eq:hat_theta_ijt} and \eqref{eq:xi_ijt}, we define
\bea
\hat{\theta}_{ij} &=& \frac{1}{n} \sum_{k=1}^{n} \left((X_{ik} - \bar X_i)(X_{jk} - \bar X_j) - \hat\sigma_{ij}\right)^2, \\
\hat \xi_{ij} &=& \frac{\hat{\theta}_{ij}}{\hat{\sigma}_{ii}\hat{\sigma}_{jj}} =  \frac{1}{n\hat \sigma_{ii}\hat \sigma_{jj}}\sum_{k=1}^{n} \left((X_{ik} - \bar X_i)(X_{jk} - \bar X_j) - \hat\sigma_{ij}\right)^2,\\
\lambda_{ij} &=& \tau\left(\frac{\log p}{n}\right)^{1/2} \left(\hat{\xi}_{ij}^{1/2} + \frac{|\hat{r}_{ij}|}{2} \left(\hat{\xi}_{ii}^{1/2} + \hat{\xi}_{jj}^{1/2} \right)\right),
\eea
where $\tau$ is the thresholding constant that can be chosen empirically through cross-validation. The correlation matrix $\R$ is then estimated by $\hat{\R}^\ast= (\hat r_{ij}^\ast)_{1\leq i,j\leq p}$ with
\begin{equation*}
\hat r_{ij}^\ast = s_{\lambda_{ij}}(\hat r_{ij}).
\end{equation*}

We consider the following class of approximately sparse correlation matrices
\begin{equation*}
\mathcal{G}^1_q(s_0(p)) = \left\{\R=(r_{ij}): \R \succ 0; \diag(\R) = 1; \max_j \sum_{i=1, i\neq j}^p |r_{ij}|^q \leq s_0(p) \right\}, \quad 0\le q <1.
\end{equation*}

The following theoretical results for $\hat\R^\ast$ can be established using a similar analysis.
\begin{proposition}\label{pr:cor_single}
Suppose $\log p = o(n^{1/3})$ and $\X$ satisfies \eqref{ineq:sub-gaussian}, \eqref{ineq:var_yy}. For $\tau > 6$, there exists some constant $C$ does not depend on $n$ or $p$ such that
\begin{equation}\label{ineq:cov_main}
\sup_{\R\in \mathcal{G}^1_q(s_0(p))} E\|\hat \R^\ast - \R\|^2 \leq C
(s_0^2(p)+1) \left(\frac{\log p}{n}\right)^{1-q} 
\end{equation}
\begin{equation}
\sup_{\R\in \mathcal{G}^1_q(s_0(p))} E\|\hat \R^\ast - \R\|_{\ell_1}^2 \leq C
(s_0^2(p)+1) \left(\frac{\log p}{n}\right)^{1-q} 
\end{equation}
\begin{equation}
\sup_{\R\in \mathcal{G}^1_q(s_0(p))} E\|\hat \R^\ast - \R\|_F^2 \leq C
p(s_0(p)+1) \left(\frac{\log p}{n}\right)^{1-q/2}.
\end{equation}
Moreover, when $\log p = o(n)$, $s_0(p) \leq Mn^{(1-q)/2}(\log p)^{-(3-q)/2}$ for some constant $M>0$, the rate in \eqref{ineq:cov_main} is optimal as we also have the lower bound
\begin{equation}
\inf_{\hat \R}\sup_{\R \in \mathcal{G}_q^1(s_0(p))} E\|\hat \R - \R \|^2 \geq cs^2_0(p) \left(\frac{\log p}{n}\right)^{1-q}
\end{equation}
\begin{equation}
\inf_{\hat \R}\sup_{\R\in \mathcal{G}^1_q(s_0(p))} E\|\hat \R - \R\|_{\ell_1}^2 \geq cs_0^2(p) \left(\frac{\log p}{n}\right)^{1-q} 
\end{equation}
\begin{equation}
\inf_{\hat \R}\sup_{\R\in \mathcal{G}^1_q(s_0(p))} E\|\hat \R - \R\|_F^2 \geq 
cps_0(p) \left(\frac{\log p}{n}\right)^{1-q/2}.
\end{equation}
\end{proposition}

\begin{remark}{\rm 
\cite{Cai_Liu} proposed an adaptive thresholding estimator $\hat \SSigma^\ast$ of a sparse covariance matrix $\SSigma$. This estimator leads naturally to an estimator $\tilde \R =(\tilde r_{ij})$  of a sparse correlation matrix $\R$  by  normalizing $\hat \SSigma^\ast = ( \hat\sigma_{ij}^\ast)$ via $\tilde r_{ij} = \hat\sigma_{ij}^\ast(\sigma_{ii}^\ast \sigma_{jj}^\ast)^{-1/2}$. The correlation matrix estimator $\tilde \R$ has similar properties as the estimator introduced above. For example, $\tilde \R$ and $\hat \R^\ast$ achieve the same rate of convergence.
 }
\end{remark}

\subsection{Optimal Estimation of Sparse Differential Covariance Matrices}

Our analysis can also be used for estimation of sparse differential covariance matrices, $\bDelta=\SSigma_1 - \SSigma_2$. Define $\theta_{ijt}$ as in \eqref{eq:theta_ijt} and its estimate $\hat\theta_{ijt}$ as in \eqref{eq:hat_theta_ijt}.
Similar to the estimation of the differential correlation matrix $\D=\R_1-\R_2$, we estimate  $\bDelta=\SSigma_1 - \SSigma_2$ by adaptive entrywise thresholding. Specifically, we define the thresholding estimator $\hat \bDelta^\ast=(\hat \delta_{ij}^\ast) \in\mathbb{R}^{p\times p}$  by
\beq
\label{Delta*}
\hat \delta^\ast_{ij} = s_{\gamma_{ij}}(\hat\sigma_{ij1} - \hat\sigma_{ij2}), \; 1\leq i,j\leq p
\eeq
where $\gamma_{ij}$ is the thresholding level given by 
\beq
\label{gamma_ij}
\gamma_{ij} = \tau\left(\left(\frac{\log p }{n_1}\hat\theta_{ij1}\right)^{1/2} + \left(\frac{\log p }{n_2}\hat\theta_{ij2}\right)^{1/2}\right).
\eeq
Same as in the last section, here $s_\lambda(z)$ belongs to the class of thresholding functions satisfying Conditions (C1)-(C3) and the thresholding constant $\tau$ can be taken chosen empirically by cross-validation.

We consider the following class of paired covariance matrices with approximately sparse differences, for $0\le q <1$,
\begin{equation}\label{eq:F} 
\mathcal{F}_q(s_0(p))
\triangleq  \left\{(\SSigma_1, \SSigma_2):  \SSigma_1, \SSigma_2 \succeq 0, \max_{1\leq i\leq p, t=1, 2} \sigma_{iit} \leq B,  \max_i \sum_{j=1}^p | \sigma_{ij1} - \sigma_{ij2}|^{q} \leq s_0(p)\right\}.
\end{equation}

Under the same conditions as those in Theorems \ref{th:main_cor} and \ref{pr:lower_bound}, a similar analysis can be used to derive the minimax upper and lower bounds. It can be shown that the estimator $\hat{\bDelta}^\ast$ given in \eqref{Delta*} with $\tau > 4$ satisfies
\begin{equation}\label{ineq:cov_diff_upper}
\sup_{(\SSigma_1, \SSigma_2)\in \mathcal{F}_q(s_0(p))} E\| \hat\bDelta^\ast - (\SSigma_1 - \SSigma_2)\|^2 \leq C (s_0^2(p)+1)\left(\frac{\log p}{n_1} + \frac{\log p}{n_2}\right)^{1-q}
\end{equation}
for some constant $C>0$.  Furthermore, the following minimax lower bound holds,
\begin{equation}\label{ineq:cov_diff_lower}
\inf_{\hat \bDelta} \sup_{(\SSigma_1, \SSigma_2) \in \mathcal{F}_q(s_0(p))} E\|\hat \bDelta - (\SSigma_1 - \SSigma_2)\|^2 \geq c s_0^2(p) \left( \frac{\log p}{n_1} + \frac{\log p}{n_2}\right)^{1-q}
\end{equation}
for some constant $c>0$. Equations \eqref{ineq:cov_diff_upper} and \eqref{ineq:cov_diff_lower} together show that the thresholding estimator $\hat \bDelta^\ast$ defined in \eqref{Delta*} and \eqref{gamma_ij} is rate-optimal.

\subsection{Estimate Differential Cross-Correlation Matrices}

In many applications such as phenome-wide association studies (PheWAS) which aims to study the relationship between a set of genomic markers $\X$ and a range of phenotypes $\Y$, the main focus is on the cross-correlations between the components of $\X$ and those of $\Y$. That is, the object of interest is a submatrix of the correlation matrix of the random vector 
$
\begin{pmatrix}
\X\\
\Y
\end{pmatrix}
$.
More specifically, let $\X=(X_{1},\ldots,X_{p_1})^{'}$ be a $p_1$-dimensional random vector and $\Y=(Y_{1},\ldots,Y_{p_2})^{'}$ be a $p_2$-dimensional random vector. In the PheWAS setting,  $\X$ may be all phenotypic disease conditions of interest and $\Y$ is a vector of genomic markers.
%

Suppose we have two independent and identical distributed samples of the $(\X, \Y)$ pairs, one for the case group and one for the control group, 
\begin{equation*}
\begin{pmatrix}
X^{(1)}_k\\
Y^{(1)}_k
\end{pmatrix}
=\begin{pmatrix}
X_{1k}^{(1)}\\
\vdots\\
X_{p_1k}^{(1)}\\
Y_{1k}\\
\vdots\\
Y_{p_2k}^{(1)}
\end{pmatrix}, \quad k = 1,\ldots, n_1;\quad \begin{pmatrix}
X^{(2)}_k\\
Y^{(2)}_k
\end{pmatrix}
=\begin{pmatrix}
X_{1k}^{(2)}\\
\vdots\\
X_{p_1k}^{(2)}\\
Y_{1k}\\
\vdots\\
Y_{p_2k}^{(2)}
\end{pmatrix}, \quad k = 1,\ldots, n_2
\end{equation*}
Here for $t = 1, 2$, $(X_k^{(t)T}, Y_k^{(t)T})^T$, $k=1,\ldots, n_1$ are independent and identical distributed samples generated from some distribution with mean $\mu_t$, covariance matrix $\SSigma_t $ and correlation matrix $\R_t$ given by
$$\mu_t = \begin{pmatrix}
\mu_{Xt}\\
\mu_{Yt}\\
\end{pmatrix},\quad \SSigma_t = \begin{bmatrix}
\SSigma_{XXt} & \SSigma_{XYt}\\
\SSigma_{YXt} & \SSigma_{YYt}
\end{bmatrix}, \quad 
\R_t = 
\begin{bmatrix}
\R_{XXt} & \R_{XYt}\\
\R_{YXt} & \R_{YYt}
\end{bmatrix}$$ 
In applications such as PheWAS, it is of special interest to estimate the differential cross-correlation matrix of $\X$ and $\Y$, i.e. $\D_{XY} = \R_{XY1} - \R_{XY2} \in \mathbb{R}^{p_1\times p_2}$. Again, we introduce the following set of paired correlation matrices with sparse cross-correlations,
\begin{equation*}
\begin{split}
\mathcal{G}_q(s_0(p_1, p_2)) = \Bigg\{&(\R_1, \R_2): \R_1, \R_2 \succeq 0, \diag(\R_1) = \diag(\R_2) = 1;\\
& \max_{1\leq i\leq p_1} \sum_{j=1}^{p_2}|(r_{XY})_{ij1} - (r_{XY})_{ij2}|^q \leq s_0(p_1, p_2)\Bigg\}, \quad 0\le q<1.
\end{split}
\end{equation*}
The thresholding procedure proposed in Section \ref{sec.procedure} can be applied to estimate $\D_{XY}$,
\begin{equation}\label{eq:D_XY}
(\hat{d}_{XY}^\ast)_{ij} = s_{\lambda_{ij}}((\hat r_{XY})_{ij1} - (\hat r_{XY})_{ij2}),\quad 1\leq i \leq p_1, \quad 1\leq j \leq p_2
\end{equation}
where $\hat\R_{XY}$ is sample cross-correlation matrix of $X$ and $Y$; $\lambda_{ij}$ is given by \eqref{eq:lambda_ij_cor}. Similar to Theorem \ref{th:main_cor},  the following theoretical results hold for  the estimator $\hat \D_{XY}^\ast= (\hat{d}_{XY}^\ast)$.
\begin{proposition}\label{pr:cross_cor}
Suppose $p = p_1+p_2$, $\log (p) = o(\min(n_1, n_2)^{1/3})$ and \eqref{ineq:sub-gaussian} and \eqref{ineq:var_yy} hold. Suppose the thresholding function $s_{\lambda}(z)$ satisfies Conditions (C1)-(C3). Then $\hat \D^\ast$ defined in \eqref{eq:D_XY} with the thresholding constant $\tau > 4$ satisfies
\begin{equation}\label{ineq:cross_bound_spe}
\sup_{(\R_1, \R_2) \in \mathcal{G}_q(s_0(p_1, p_2))} E\|\hat\D_{XY}^\ast - (\R_{XY1} -\R_{XY2})\|^2 \leq  C(s_0^2(p)+1) \left(\frac{\log p}{n_1} + \frac{\log p}{n_2}\right)^{1-q}
\end{equation}
\begin{equation}\label{ineq:cross_bound_l1}
\sup_{(\R_1, \R_2) \in \mathcal{G}_q(s_0(p_1, p_2))} E\|(\hat\D_{XY}^\ast - (\R_{XY1} -\R_{XY2}))^\T\|_{\ell_1}^2 \leq C
(s_0^2(p)+1) \left(\frac{\log p}{n_1} + \frac{\log p}{n_2}\right)^{1-q}
\end{equation}
\begin{equation}\label{ineq:cross_bound_F}
\sup_{(\R_1, \R_2) \in \mathcal{G}_q(s_0(p_1, p_2))} E\|\hat\D_{XY}^\ast - (\R_{XY1} -\R_{XY2})\|_F^2 \leq Cp(s_0(p)+1) \left(\frac{\log p}{n_1} + \frac{\log p}{n_2}\right)^{1-q/2}
\end{equation}
for some constant $C>0$ that does not depend on $n_1, n_2$ or $p$.
\end{proposition}

The proof of Proposition \ref{pr:cross_cor} is similar to that of Theorem \ref{th:main_cor} by analyzing the block $\hat{\D}_{XY} - (\R_{XY1} - \R_{XY2})$ instead of the whole matrix $\D^\ast - (\R_1 - \R_2)$. We omit the detailed proof here.


\bibliographystyle{apa}
\bibliography{reference_add}

\section*{Appendix: Proofs}

We prove the main theorems in the Appendix. Throughout the Appendix, we denote by $C$ a constant which does not depend on $p, n_1$ and $n_2$, and may vary from place to place.

{\noindent\bf Proof of Theorem~\ref{th:main_cor}} To prove this theorem, we consider the following three events separately,
\bea\label{eq:A_1}
A_1 &=& \left\{\max_{ijt}\frac{|\hat{\sigma}_{ijt} - \sigma_{ijt}|}{\left( \hat{\theta}_{ijt}\log p/n_t\right)^{1/2}} \leq \frac{\tau}{4}+3 , \quad \text{and}\quad \max_{ijt} \frac{|\hat{\theta}_{ijt} - \theta_{ijt}|}{\sigma_{iit}\sigma_{jjt}} \leq \varepsilon \right\}
\\
A_2 &=&\left\{ \max_{ijt} \frac{|\hat{\sigma}_{ijt} - \sigma_{ijt}|}{\left( \hat{\theta}_{ijt}\log p/n_t\right)^{1/2}} > \frac{\tau}{4}+3 , \quad \max_{ijt}\frac{|\hat{\theta}_{ijt} - \theta_{ijt}|}{\sigma_{iit}\sigma_{jjt}} \leq \varepsilon,\right. \nonumber \\
&&\quad \quad \text{and}\quad \left. \max_{ijt}\frac{|\hat{\sigma}_{ijt} - \sigma_{ijt}|}{(\sigma_{iit}\sigma_{jjt})^{1/2}} \leq \min(0.5, C_1C_3) \right\} \label{eq:A_2}\\
A_3 &=& (A_1\cup A_2)^c. \label{eq:A_3}
\eea

Here $\varepsilon$ is the fixed constant which satisfies $0<\varepsilon < \nu_0/2$ where $\nu_0$ is introduce in \eqref{ineq:var_yy}; $C_1$ and $C_3$ are constants which do not depends on $p, n_1, n_2$ and shall be specified later in Lemma \ref{lm:Cai_Liu_lemma2}. 

\begin{enumerate}
\item First we would like to show that under the event $A_1$,
\begin{equation}\label{ineq:A_1_spe}
\|\hat{\D}^\ast - (\R_1 - \R_2)\|^2 \leq Cs_0^2(p) \left(\frac{\log p}{n_1} + \frac{\log p}{n_2}\right)^{1-q},
\end{equation}
\begin{equation}\label{ineq:A_1_l1}
\|\hat{\D}^\ast - (\R_1 - \R_2)\|_{\ell_1}^2 \leq Cs_0^2(p) \left(\frac{\log p}{n_1} + \frac{\log p}{n_2}\right)^{1-q},
\end{equation}
\begin{equation}\label{ineq:A_1_F}
\|\hat{\D}^\ast - (\R_1 - \R_2)\|_F^2 \leq Cps_0(p) \left(\frac{\log p}{n_1} + \frac{\log p}{n_2}\right)^{1-q/2}.
\end{equation}
In fact,
$$EY_i^{(t)4} \leq \frac{2E\exp(\eta Y_i^{(t)2})}{\eta^2} \leq \frac{2K}{ \eta^2}$$
for all $1\leq i \leq p$, so
\begin{equation*}
\begin{split}
\theta_{ijt} = & {\rm Var}(X_i^{(t)} - \mu_i)(X_j^{(t)} - \mu_j) \leq E(X_i^{(t)} - \mu_i)^2 (X_j^{(t)} - \mu_j)^2\\
\leq & \left(E(X_i^{(t)} - \mu_i)^4E(X_j^{(t)} - \mu_j)^4\right)^{1/2} = \sigma_{iit}\sigma_{jjt}\left(EY_i^{(t)4}EY_j^{(t)4}\right)^{1/2} \leq C\sigma_{iit}\sigma_{jjt},
\end{split}
\end{equation*}
\begin{equation}\label{ineq:theta>=tau_sigma}
\theta_{ijt} = {\rm Var}(Y_iY_j)\cdot \sigma_{iit}\sigma_{jjt} \overset{\eqref{ineq:var_yy}}{\geq} \nu_0\sigma_{iit}\sigma_{jjt}.
\end{equation}
So by the definition of $A_1$, we have
\begin{equation}\label{ineq:theta<=sigmasigma}
\hat\theta_{ijt} \leq \theta_{ijt} + |\hat\theta_{ijt} -\theta_{ijt}| \leq C\sigma_{iit}\sigma_{jjt}, \text{ for all } i, j, t,
\end{equation}
\begin{equation}
\hat{\theta}_{ijt} \geq \theta_{ijt} - |\hat\theta_{ijt} - \theta_{ijt}| \geq \nu_0\sigma_{iit}\sigma_{jjt} - \varepsilon \sigma_{iit}\sigma_{jjt} \geq \frac{\nu_0}{2}\sigma_{iit}\sigma_{jjt}.
\end{equation}
Hence, 
\begin{equation}\label{ineq:hat_sigma/sigma}
\left|\frac{\hat{\sigma}_{iit}}{\sigma_{iit}} - 1\right| \leq \frac{|\hat{\sigma}_{iit} - \sigma_{iit}|}{\sigma_{iit}}  \overset{\eqref{eq:A_1}}{\leq} \frac{\tau/4+3}{\sigma_{iit}}\left(\log p \frac{\hat{\theta}_{iit}}{n_t}\right)^{1/2}  \overset{\eqref{ineq:theta<=sigmasigma}}{\leq} C\left(\frac{\log p}{n_t}\right)^{1/2} 
\end{equation}
\begin{equation}\label{ineq:sigma/hat_sigma}
\left|\frac{\sigma_{iit}}{\hat\sigma_{iit}} - 1\right| \leq \frac{|\hat{\sigma}_{iit} - \sigma_{iit}|}{\hat\sigma_{iit}}  \overset{\eqref{eq:A_1}}{\leq} (\tau/4+3)\left(\frac{\log p}{n_t}\right)^{1/2} \frac{\hat{\theta}_{iit}^{1/2}}{\hat\sigma_{iit}}
\end{equation}
Suppose $x = \sigma_{iit}/\hat{\sigma}_{iit}$, $y = \sigma_{jjt}/\hat{\sigma}_{jjt}$. By \eqref{ineq:hat_sigma/sigma} and $\left(\frac{\log p}{n_t}\right)^{1/2} \to 0$, we have $\max \left\{|x - 1|, |y-1|\right\} \leq C\left(\frac{\log p}{n_t}\right)^{1/2}$ when $n_t$ is large enough. Thus for large $n_t$, we obtain
\begin{equation}\label{ineq:sqrt_xy-1}
\begin{split}
\left|\left(\frac{\sigma_{iit}\sigma_{jjt}}{\hat{\sigma}_{iit}\hat{\sigma}_{jjt}}\right)^{1/2}-1\right|= & \left|(xy)^{1/2} - 1\right| = \frac{|xy - 1|}{(xy)^{1/2} + 1} \leq \frac{|x-1| + x|y-1|}{2 - \max\left(|x-1|, |y-1|\right)}\\
\leq & \frac{\max(1, x)}{2-\max(|x-1|, |y-1|)}\left(|x-1|+|y-1|\right)\\
\leq & \left(\frac{1}{2} + O\left(\left(\frac{\log p}{n_t}\right)^{1/2}\right)\right) \left(|x-1| + |y-1|\right).
\end{split}
\end{equation}
It then follows from the assumption $\log p = o(n_t^{1/3})$ that for large $n_t$, 
\begin{equation}\label{ineq:xi_bound2}
\hat \xi_{ijt} = \frac{\hat{\theta}_{ijt}}{\hat\sigma_{iit}\hat\sigma_{jjt}} \overset{\eqref{ineq:theta<=sigmasigma}}{\leq} \frac{C\sigma_{iit}\sigma_{jjt}}{\hat{\sigma}_{iit}\hat \sigma_{jjt}} \overset{\eqref{ineq:hat_sigma/sigma}}{\leq} C
\end{equation}
and
\begin{equation*}
\begin{split}
|\hat r_{ijt} - r_{ijt}|  = &  \left|\frac{\hat \sigma_{ijt}}{(\hat\sigma_{iit}\hat\sigma_{jjt})^{1/2}} -  \frac{ \sigma_{ijt}}{(\sigma_{iit}\sigma_{jjt})^{1/2}}\right| \leq \frac{|\hat \sigma_{ijt} - {\sigma}_{ijt}|}{(\hat\sigma_{iit}\hat\sigma_{jjt})^{1/2}} + \frac{|\sigma_{ijt}|}{(\sigma_{iit}\sigma_{jjt})^{1/2}}\left|\left(\frac{\sigma_{iit}\sigma_{jjt}}{\hat\sigma_{iit}\hat\sigma_{jjt}}\right)^{1/2} - 1\right|\\
 \overset{\eqref{eq:A_1}\eqref{ineq:sqrt_xy-1}}{\leq} & \left(\frac{\tau}{4} + 3\right)\sqrtp{\frac{\log p}{n_t} \frac{\hat{\theta}_{ijt}}{\hat{\sigma}_{iit}\hat{\sigma}_{jjt}}} + |r_{ijt}|\left(\frac{1}{2} + O\left(\sqrtp{\frac{\log p}{n_t}}\right)\right)\left( \left|\frac{\sigma_{iit}}{\hat \sigma_{iit}} - 1\right| + \left|\frac{\sigma_{jjt}}{\hat \sigma_{jjt}} - 1\right|\right)\\
 \overset{\eqref{ineq:sigma/hat_sigma}}{\leq} & \left(\frac{\tau}{4}+3\right)\sqrtp{\frac{\log p}{n_t}}\left(\sqrtp{\frac{\hat{\theta}_{ijt}}{\hat{\sigma}_{iit}\hat{\sigma}_{jjt}}} + |r_{ijt}|\left(\frac{1}{2} + O\left(\sqrtp{\frac{\log p}{n_t}}\right)\right)\left(\frac{\hat{\theta}_{iit}^{1/2}}{\hat \sigma_{iit}} + \frac{\hat{\theta}_{jjt}^{1/2}}{\hat \sigma_{jjt}}\right)\right)\\
 \leq & \left(\frac{\tau}{2}+2\right)\sqrtp{\frac{\log p}{n_t}}\left(\hat\xi_{ijt}^{1/2} + \frac{|r_{ijt}|}{2}\left(\hat{\xi}_{iit}^{1/2} + \hat{\xi}_{jjt}^{1/2}\right)\right)\\
 \leq & \left(\frac{\tau}{2}+2\right)\sqrtp{\frac{\log p}{n_t}} \left(\hat\xi_{ijt}^{1/2} + \frac{|\hat r_{ijt}|}{2}\left(\hat{\xi}_{iit}^{1/2} + \hat{\xi}_{jjt}^{1/2}\right)\right)\\
& + |\hat r_{ijt} - r_{ijt}|\left(\frac{\tau}{4}+1\right)\sqrtp{\frac{\log p}{n_t}}\left(\hat{\xi}_{iit}^{1/2} + \hat{\xi}_{jjt}^{1/2}\right)\\
\overset{\eqref{ineq:xi_bound2}}{\leq} &\left(\frac{\tau}{2}+2\right)\sqrtp{\frac{\log p}{n_t}} \left(\hat\xi_{ijt}^{1/2} + \frac{|\hat r_{ijt}|}{2}\left(\hat{\xi}_{iit}^{1/2} + \hat{\xi}_{jjt}^{1/2}\right)\right) + C\sqrtp{\frac{\log p}{n_t}}|\hat r_{ijt} - r_{ijt}|.
\end{split}
\end{equation*}
We shall note the difference between $\frac{|r_{ijt}|}{2}$ and $\frac{|\hat r_{ijt}|}{2}$ above. Next, we rearrange the inequality above and write it into an inequality for $|\hat r_{ijt} - r_{ijt}|$,
\begin{equation}\label{ineq:hat r - r final}
\begin{split}
|\hat r_{ijt} - r_{ijt}| \leq & \frac{\left(\frac{\tau}{2}+2\right)\sqrtp{\frac{\log p}{n_t}} \left(\hat\xi_{ijt}^{1/2} + \frac{|\hat r_{ijt}|}{2}\left(\hat{\xi}_{iit}^{1/2} + \hat{\xi}_{jjt}^{1/2}\right)\right)}{1 - C\sqrtp{\frac{\log p}{n_t}}}\\
 \leq & \tau\sqrtp{\frac{\log p}{n_t}} \left( \hat\xi_{ijt}^{1/2} + \frac{|\hat r_{ijt}|}{2}\left(\hat\xi_{iit}^{1/2} + \hat\xi_{jjt}^{1/2}\right) \right) \overset{\eqref{eq:lambda_ij_cor}}{=} \lambda_{ijt}.
\end{split}
\end{equation}
\eqref{ineq:hat r - r final} implies 
\begin{equation}\label{ineq:hat_r - r}
|(\hat r_{ij1} - \hat r_{ij2}) - (r_{ij1} - r_{ij2})| \leq \lambda_{ij1} + \lambda_{ij2} = \lambda_{ij} \text{  holds for all  } 1\leq i, j\leq p
\end{equation}
Next, by \eqref{ineq:hat_r - r} and (C1) and (C3) of $s_\lambda(z)$, 
\begin{equation}
\begin{split}
\left|s_{\lambda_{ij}} (\hat r_{ij1} - \hat r_{ij2}) - (r_{ij1} - r_{ij2})\right| & \leq \left|s_{\lambda_{ij}} (\hat r_{ij1} - \hat r_{ij2})\right| + \left|(r_{ij1} - r_{ij2})\right| \\
& \leq (1 +c)|r_{ij1} - r_{ij2}|, 
\end{split}
\end{equation}
\begin{equation}
\begin{split}
& \left|s_{\lambda_{ij}} (\hat r_{ij1} - \hat r_{ij2}) - (r_{ij1} - r_{ij2})\right|\\
 \leq &  \left|s_{\lambda_{ij}} (\hat r_{ij1} - \hat r_{ij2}) - (\hat r_{ij1} - \hat r_{ij2})\right| + \left|(\hat r_{ij1} - \hat r_{ij2}) - (r_{ij1} - r_{ij2})\right| 
  \leq  2\lambda_{ij}, 
\end{split}
\end{equation}
which implies
\begin{equation}\label{ineq:main_middle_l1}
\left| s_{r_{ij}}(\hat r_{ij1} - \hat r_{ij2}) - (r_{ij1} - r_{ij2})\right| \leq 
(2\lambda_{ij})^{1-q}(1+c)^q| r_{ij1} - r_{ij2}|^q,
\end{equation}
\begin{equation}\label{ineq:main_middle_F}
\left| s_{r_{ij}}(\hat r_{ij1} - \hat r_{ij2}) - (r_{ij1} - r_{ij2})\right|^2 \leq 
(2\lambda_{ij})^{2-q}(1+c)^q| r_{ij1} - r_{ij2}|^q,
\end{equation}
where $0\leq q< 1$. Hence, 
\begin{equation*}
\begin{split}
& \|\hat{\D}^\ast - (\R_1 - \R_2)\|_{\ell_1}\\
= & \max_{i} \sum_{j=1}^p \left|s_{\lambda_{ij}}(\hat r_{ij1} - \hat r_{ij2}) - (r_{ij1} - r_{ij2})\right|\\
  \overset{\eqref{ineq:main_middle_l1}}{\leq} & \max_i 2^{1-q}(1+c)^q\sum_{j=1}^p \lambda_{ij}^{1-q}|r_{ij1} - r_{ij2}|^q\\
 \overset{\eqref{eq:lambda_ij_cor}}{\leq} & \max_i 2^{1-q}(1+c)^q\sum_{j=1}^p \Big\{\tau^{1-q}(\log p)^{(1-q)/2}\\
 & \times \left(\frac{\hat\xi_{ij1}^{1/2} + |\hat r_{ij1}|(\hat\xi_{ii1}^{1/2} + \hat\xi_{jj1}^{1/2})/2}{n_1^{1/2}} + \frac{\hat\xi_{ij2}^{1/2} + |\hat r_{ij2}|(\hat\xi_{ii2}^{1/2} + \hat\xi_{jj2}^{1/2})/2}{n_2^{1/2}}\right)^{1-q}|r_{ij1} - r_{ij2}|^q\Big\} \\
 \overset{\eqref{ineq:xi_bound2}}{\leq} & \max_i C \sum_{j=1}^p\left\{(\log p)^{(1-q)/2} \left(\frac{1}{n_1} + \frac{1}{n_2}\right)^{(1-q)/2}|r_{ij1}-r_{ij2}|^{q}\right\}\\
 \leq & Cs_0(p) \left(\frac{\log p}{n_1} + \frac{\log p}{n_2}\right)^{(1-q)/2}.
\end{split}
\end{equation*}
which yields to \eqref{ineq:A_1_l1}. \eqref{ineq:A_1_spe} also holds due to the fact that $\|A\|_2\leq \|A\|_{L_1}$ for any symmetric matrix $A$. 
Similarly,
\begin{equation*}
\begin{split}
&\left\|\hat \D^\ast - (\R_1 - \R_2)\right\|_F^2\\ 
= & \sum_{i=1}^p\sum_{j=1}^p \left|s_{\lambda_{ij}}(\hat r_{ij1} - \hat r_{ij2}) - (r_{ij1} - r_{ij2})\right|^2\\
  \overset{\eqref{ineq:main_middle_F}}{\leq} & 2^{2-q}(1+c)^q\sum_{i=1}^p\sum_{j=1}^p \lambda_{ij}^{2-q}|r_{ij1} - r_{ij2}|^q\\
 \overset{\eqref{eq:lambda_ij_cor}}{\leq} & 2^{2-q}(1+c)^q\sum_{i=1}^p\sum_{j=1}^p \Bigg\{\tau^{2-q}(\log p)^{(2-q)/2}\\
 & \times\left(\frac{\hat\xi_{ij1}^{1/2} + |\hat r_{ij1}|(\hat\xi_{ii1}^{1/2} + \hat\xi_{jj1}^{1/2})/2}{n_1^{1/2}} + \frac{\hat\xi_{ij2}^{1/2} + |\hat r_{ij2}|(\hat\xi_{ii2}^{1/2} + \hat\xi_{jj2}^{1/2})/2}{n_2^{1/2}}\right)^{2-q}|r_{ij1} - r_{ij2}|^q\Bigg\} \\
 \overset{\eqref{ineq:xi_bound2}}{\leq} & Cp \max_i \sum_{j=1}^p\left\{(\log p)^{(2-q)/2} \left(\frac{1}{n_1} + \frac{1}{n_2}\right)^{(2-q)/2}|r_{ij1}-r_{ij2}|^{q}\right\}\\
 \leq & Cps_0(p) \left(\frac{\log p}{n_1} + \frac{\log p}{n_2}\right)^{1-q/2}.
\end{split}
\end{equation*}
which implies \eqref{ineq:A_1_F}.

\item For $A_2$, we wish to prove,
\begin{equation}\label{ineq:A_2_spe}
\int_{A_2} \| \hat{\D}^\ast - (\R_1 - \R_2)\|^2 dP \leq C(p^{-\tau/4+1}\log p)\left(\frac{1}{n_1} + \frac{1}{n_2}\right)
\end{equation}
\begin{equation}\label{ineq:A_2_l1}
\int_{A_2} \| \hat{\D}^\ast - (\R_1 - \R_2)\|_{\ell_1}^2 dP \leq C(p^{-\tau/4+1}\log p)\left(\frac{1}{n_1} + \frac{1}{n_2}\right)
\end{equation}
\begin{equation}\label{ineq:A_2_F}
\int_{A_2} \| \hat{\D}^\ast - (\R_1 - \R_2)\|_F^2 dP \leq C(p^{-\tau/4+1}\log p)\left(\frac{1}{n_1} + \frac{1}{n_2}\right)
\end{equation}
In order to prove these probability bounds, we introduce the following lemma, which revealed the relationship between $\hat\theta_{ijt}, \theta_{ijt}$ and $\hat{\sigma}_{ijt}$, $\sigma_{ijt}$.
\begin{lemma} \label{lm:Cai_Liu_lemma2}
For any $\tau> 0$,
\begin{equation}\label{ineq:main_condition1}
{\rm pr}\left(\max_{i, j, t}|\hat\sigma_{ijt} - \sigma_{ijt}| > (\tau/4 + 3) \sqrtp{\hat\theta_{ijt}\log p / n_t}\right) \leq C(\log p)^{-1/2}p^{-\tau/4-1};
\end{equation}
There exist constants $C_1, C_2, C_3$ which do not depend on $p, n_1, n_2$ such that
\begin{equation}\label{ineq:main_condition11}
\begin{split}
{\rm pr}\left(\max_{i,j} \frac{|\hat{\sigma}_{ijt} - \sigma_{ijt}|}{\sqrtp{\sigma_{iit}\sigma_{jjt}}} > C_1x\right) \leq C_2p^2\left(\exp(-n_tx^2)\right), \quad \text{for all }0< x \leq C_3, t = 1,2;
\end{split}
\end{equation}
For any $\varepsilon>0$ and $M>0$,
\begin{equation}\label{ineq:main_condition2}
{\rm pr}\left(\max_{i, j, t}|\hat{\theta}_{ijt} - \theta_{ijt}| / (\sigma_{iit}\sigma_{jjt}) > \varepsilon \right) \leq Cp^{-M}\left(1/n_1 + 1/n_2\right)
\end{equation}
\end{lemma}
The proof of Lemma \ref{lm:Cai_Liu_lemma2} is given later. Note that \eqref{ineq:main_condition1} immediately leads to 
\begin{equation}\label{ineq:P_A_2}
{\rm pr}(A_2) \leq C(\log p)^{-1/2} p^{-\tau/4-1}.
\end{equation} 
By the definition of $A_2$ \eqref{eq:A_2}, we still have \eqref{ineq:theta<=sigmasigma}. Besides, by the definition of $A_2$, $\frac{|\hat\sigma_{iit} - \sigma_{iit}|}{\sigma_{iit}} \leq 0.5$, which leads to $\hat \sigma_{iit} \geq 0.5\sigma_{iit}$. Thus,
\begin{equation}\label{ineq:A2_hat_xi}
\hat\xi_{ijt} = \frac{\hat{\theta}_{ijt}}{\hat\sigma_{iit}\hat\sigma_{jjt}} \leq \frac{C \sigma_{iit}\sigma_{jjt}}{(0.5 \sigma_{iit})(0.5\sigma_{jjt})} = 4C.
\end{equation}

For convenience, we denote the random variable
\begin{equation}
T = \max_{ijt} \frac{|\hat{\sigma}_{ijt} - \sigma_{ijt}|}{\sqrtp{\sigma_{iit}\sigma_{jjt}}}.
\end{equation}
Under $A_2$, we have $T \leq 0.5$. Then for all $1\leq i,j\leq p, t =1,2$,
\begin{equation*}
\begin{split}
\hat r_{ijt} - r_{ijt} & = \frac{\hat\sigma_{ijt}}{\sqrtp{\hat\sigma_{iit}\hat\sigma_{jjt}}} - \frac{\sigma_{ijt}}{\sqrtp{\sigma_{iit}\sigma_{jjt}}}\\
& = \frac{\frac{\hat{\sigma}_{ijt}}{\sqrtp{\sigma_{iit}\sigma_{jjt}}} }{\sqrtp{\hat{\sigma}_{iit}/\sigma_{iit}} \times \sqrtp{\hat{\sigma}_{jjt}/\sigma_{jjt}}} - \frac{\sigma_{ijt}}{\sqrtp{\sigma_{iit}\sigma_{jjt}}}\\
& \leq \frac{\frac{\sigma_{ijt}}{\sqrtp{\sigma_{iit}\sigma_{jjt}}} + T }{\sqrtp{\sigma_{iit}/\sigma_{iit} - T} \times \sqrtp{\sigma_{jjt}/\sigma_{jjt} - T} } - \frac{\sigma_{ijt}}{\sqrtp{\sigma_{iit}\sigma_{jjt}} }\\
& = \frac{r_{ijt} + T}{1-T} - r_{ijt}\\
& \leq (1 + 2T)(r_{ijt} + T) - r_{ijt}\\
& \leq 4T.
\end{split}
\end{equation*}
Similarly calculation also leads to $\hat r_{ijt} - r_{ijt} \geq -4T$. Then, by (C3) of $s_{\lambda_{ij}}(z)$,
\begin{equation}\label{ineq:A2_l1_bound}
\begin{split}
& \|\hat \D^\ast - (\R_1 - \R_2)\|_{\ell_1}^2= \max_i \left(\sum_{j=1}^p |s_{\lambda_{ij}}(\hat r_{ij1} - \hat r_{ij2}) - (r_{ij1} - r_{ij2})|\right)^2\\
\leq & \max_i \left(\sum_{j=1}^p\left(|s_{\lambda_{ij}}(\hat r_{ij1} - \hat r_{ij2}) - (\hat r_{ij1} - \hat r_{ij2})| + |(\hat r_{ij1} - \hat r_{ij2}) - (r_{ij1} - r_{ij2})|\right)\right)^2\\
\leq & \max_i\left(\sum_{j=1}^p \left(\lambda_{ij} + 8T\right)\right)^2
\overset{\eqref{eq:lambda_ij_cor}\eqref{ineq:A2_hat_xi}}{\leq} Cp^2\left(\frac{\log p}{n_1} + \frac{\log p}{n_2}+T^2\right).
\end{split}
\end{equation}
In addition, due to $\|\cdot\|_{\ell_1} \geq \|\cdot\|$, we also have $\|\hat{\D}^\ast - (\R_1 - \R_2)\|^2 \leq Cp^2\left(\frac{\log p}{n_1} + \frac{\log p}{n_2} + T^2\right).$
Similarly,
\begin{equation}
\begin{split}
& \|\hat{\D}^\ast - (\R_1 - \R_2) \|_F^2\\
 = & \sum_{i=1}^p\sum_{j=1}^p |s_{\lambda_{ij}}(\hat r_{ij1} - \hat r_{ij2}) - (r_{ij1} - r_{ij2})|^2\\
\leq & 2\sum_{i=1}^p\sum_{j=1}^p\left( |s_{\lambda_{ij}}(\hat r_{ij1} - \hat r_{ij2}) - (\hat r_{ij1} - \hat r_{ij2})|^2 +|(\hat r_{ij1} - \hat r_{ij2}) - (r_{ij1} - r_{ij2})|^2\right)\\
\leq & Cp^2\left(\frac{\log p}{n_1} + \frac{\log p}{n_2} + T^2\right)
\end{split}
\end{equation}
Therefore,
\begin{equation}
\begin{split}
& \int_{A_2}\|\hat{\D}^\ast - (\R_1 - \R_2)\|^2_{\ell_1} dP\\
\overset{\eqref{ineq:A2_l1_bound}}{\leq} & \int_{A_2} Cp^2\left(\frac{\log p}{n_1} + \frac{\log p}{n_2} + T^2\right)dP\\
\leq & Cp^2\left(\frac{\log p}{n_1}+\frac{\log p}{n_2}\right) {\rm pr}(A_2) + Cp^2\int_0^{\min(0.5, C_1C_3)}2x {\rm pr}(\{T \geq x\}\cap A_2)dx\\
\leq & Cp^2\left(\frac{\log p}{n_1} + \frac{\log p}{n_2}\right){\rm pr}(A_2) + Cp^2 \int_0^{C_1\sqrtp{M\log p(1/n_1+1/n_2)}}2x{\rm pr}(A_2)dx\\
& + \int_{C_1\sqrtp{M\log p(1/n_1+1/n_2)}}^{\min(0.5, C_1C_3)}2x{\rm pr}(T\geq x)dx\\
\overset{\eqref{ineq:main_condition11}}{\leq} & Cp^2\left(\frac{\log p}{n_1} + \frac{\log p}{n_2}\right){\rm pr}(A_2) + Cp^2\left(\frac{\log p}{n_1} + \frac{\log p}{n_2}\right){\rm pr}(A_2)\\
& + \int_{C_1\sqrtp{M\log p(1/n_1+1/n_2)}}^{+\infty}2xC_2\left(\exp(-n_1(x/C_1)^2) + \exp(-n_2(x/C_1)^2)\right)dx\\
\leq & Cp^2\left(\frac{\log p}{n_1} + \frac{\log p}{n_2}\right){\rm pr}(A_2)\\
 & + C\left(\frac{1}{n_1}\exp(-n_1(x/C_1)^2) + \frac{1}{n_2}\exp(-n_2(x/C_1)^2)\right)\Big|_{+\infty}^{C_1\sqrtp{M\log p(1/n_1+1/n_2)}}\\
\overset{\eqref{ineq:P_A_2}}{\leq} & Cp^{-\tau/4+1}\left(\frac{\log p}{n_1} + \frac{\log p}{n_2}\right) + Cp^{-M}\left(\frac{1}{n_1} + \frac{1}{n_2}\right)\\
\end{split}
\end{equation}
Similarly, we have
$$\int_{A_2}\|\hat{\D}^\ast - (\R_1 - \R_2)\|^2 dP \leq Cp^{-\tau/4+1}\left(\frac{\log p}{n_1} + \frac{\log p}{n_2}\right) + Cp^{-M}\left(\frac{1}{n_1} + \frac{1}{n_2}\right), $$
$$\int_{A_2}\|\hat{\D}^\ast - (\R_1 - \R_2)\|^2_F dP \leq Cp^{-\tau/4+1}\left(\frac{\log p}{n_1} + \frac{\log p}{n_2}\right) + Cp^{-M}\left(\frac{1}{n_1} + \frac{1}{n_2}\right), $$
which finishes the proof of \eqref{ineq:A_2_spe}, \eqref{ineq:A_2_l1} and \eqref{ineq:A_2_F} when we choose $M>\tau/4-1$.

\item For $A_3$, \eqref{ineq:main_condition2} and $\log p =o(n^{1/3})$ leads to 
\begin{equation}\label{ineq:A_3_P}
\begin{split}
{\rm pr}(A_3) \leq & {\rm pr}\left(\max_{ijt} \frac{|\hat{\theta}_{ijt} - \theta_{ijt}|}{\sigma_{iit}\sigma_{jjt}} > \varepsilon\right) + {\rm pr}\left(\max_{ijt}\frac{|\hat{\sigma}_{ijt} - \sigma_{ijt}|}{\sqrtp{\sigma_{iit}\sigma_{jjt}}} > \min(0.5, C_1C_3)\right)\\
\leq & Cp^{-M}(1/n_1+1/n_2) + C_2p^2\left(\exp(-n_1 \min(\frac{1}{2C_1},C_3)^2) + \exp(-n_2 \min(\frac{1}{2C_1}, C_3)^2)\right)\\
= & Cp^{-M}(1/n_1+1/n_2)
\end{split}
\end{equation} 
Besides, since $r_{ijt}, \hat r_{ijt}$ are the population and sample correlations, $|r_{ijt}|\leq1$, $|\hat r_{ijt}|\leq 1$. By (C1) of thresholding $s_{\lambda}(z)$, we have $|s_{\lambda}(x) - x| \leq c|x|$ for all $x\in \mathbb{R}$. Thus,
\begin{equation*}
\begin{split}
|s_{\lambda_{ij}}(\hat r_{ij1} - \hat r_{ij2}) - (r_{ij1} - r_{ij2})|\leq & |r_{ij1}| + |r_{ij2}| + |s_{\lambda_{ij}}(\hat r_{ij1} - \hat r_{ij2})|\\
\leq & 2 + c|\hat r_{ij1} -\hat r_{ij2}|\leq 2 + 2c
\end{split}
\end{equation*}
which yields
\begin{equation}
\|\hat{\D}^\ast - (\R_1 - \R_2)\|_{\ell_1}^2 = \max_i \left(\sum_{j=1}^p |s_{\lambda_{ij}}(\hat r_{ij1} - \hat r_{ij2}) - (r_{ij1} - r_{ij2})|\right)^2 \leq (2+2c)^2p^2
\end{equation}
Similarly, $\|\hat{\D}^\ast - (\R_1 - \R_2)\|^2 \leq (2+2c)^2p^2$, 
$\|\hat{\D}^\ast - (\R_1 - \R_2)\|_F^2 \leq (2+2c)^2p^2$.
Therefore,
\begin{equation}\label{ineq:A_3_spe}
\int_{A_3} \| \hat{\D}^\ast - (\R_1 - \R_2)\|^2 dP \overset{\eqref{ineq:A_3_P}}{\leq} Cp^{-M+2}\left(\frac{1}{n_1} + \frac{1}{n_2}\right)
\end{equation}
\begin{equation}\label{ineq:A_3_l1}
\int_{A_3} \| \hat{\D}^\ast - (\R_1 - \R_2)\|_{\ell_1}^2 dP \overset{\eqref{ineq:A_3_P}}{\leq} Cp^{-M+2}\left(\frac{1}{n_1} + \frac{1}{n_2}\right)
\end{equation}
\begin{equation}\label{ineq:A_3_F}
\int_{A_3} \| \hat{\D}^\ast - (\R_1 - \R_2)\|_F^2 dP \overset{\eqref{ineq:A_3_P}}{\leq} Cp^{-M+2}\left(\frac{1}{n_1} + \frac{1}{n_2}\right)
\end{equation}

\end{enumerate}

Finally, we combine the situations of $A_1, A_2$ and $A_3$. When $\tau > 4$ and $M >2$, we have
\begin{equation}
\begin{split}
E\|\hat{\D}^\ast - (\R_1 - \R_2)\|^2  = & \left(\int_{A_1} + \int_{A_2} + \int_{A_3} \right)\| \hat{\D}^\ast - (\R_1 - \R_2)\|^2dP \\
\overset{\eqref{ineq:A_1_spe}\eqref{ineq:A_2_spe}\eqref{ineq:A_3_spe}}{\leq} & C\left(s_0^2(p) + 1\right)\left(\frac{\log p}{n_1} + \frac{\log p}{n_2}\right)^{1-q}
\end{split}
\end{equation}
which has proved \eqref{ineq:main_bound_spe}. \eqref{ineq:main_bound_l1} and \eqref{ineq:main_bound_F} can be proved similarly by \eqref{ineq:A_1_l1}, \eqref{ineq:A_2_l1}, \eqref{ineq:A_3_l1} and \eqref{ineq:A_1_F}, \eqref{ineq:A_2_F}, \eqref{ineq:A_3_F}.
\quad $\square$

{\noindent\bf Proof of Lemma~\ref{lm:Cai_Liu_lemma2}.} \eqref{ineq:main_condition1} is directly from (25) in \cite{Cai_Liu}. For \eqref{ineq:main_condition2}, the proof is essentially the same as the proof of (26) in \cite{Cai_Liu} as long as we use $x = ((M+2)\log p + \log n)^{1/2}$ in stead of $x = ((M+2)\log p)^{1/2}$ in their proof. Now we mainly focus on the proof of \eqref{ineq:main_condition11}. Without loss of generality, we can translate $X$ and assume that $\m_1 = \m_2 =0$. Note that we have the following formulation,
\begin{equation}\label{eq:hatsigma_ijt - sigma_ijt}
\frac{\hat{\sigma}_{ijt} - \sigma_{ijt}}{\sqrtp{\sigma_{iit}\sigma_{jjt}}} = \frac{1}{n_t}\sum_{k=1}^{n_t}\frac{(X_{ik}^{(t)}X_{jk}^{(t)} - \sigma_{ijt})}{\sqrtp{\sigma_{iit}\sigma_{jjt}}} - \frac{\bar X_i^{(t)} \bar X_j^{(t)}}{\sqrtp{\sigma_{iit}\sigma_{jjt}}} = \left(\frac{1}{n_t}\sum_{k=1}^{n_t} (Y_{ik}^{(t)}Y_{jk}^{(t)} - r_{ijt}) - \bar Y_i^{(t)} \bar Y_j^{(t)}\right) 
\end{equation}
Since 
\begin{equation*}
\begin{split}
& E (Y_{i}^{(t)}Y_{j}^{(t)} - r_{ijt})^2 e^{\frac{\eta}{2}|Y_{i}^{(t)}Y_{j}^{(t)} - r_{ijt}|}\\
 \leq & \frac{4}{\eta^2}Ee^{\eta |Y_{i}^{(t)}Y_{j}^{(t)} - r_{ijt}|} \leq \frac{4}{\eta^2}Ee^{\eta (Y_{i}^{(t)}Y_{j}^{(t)} - r_{ijt})} + \frac{4}{\eta^2}Ee^{-\eta(Y_{i}^{(t)}Y_{j}^{(t)}-r_{ijt})}\\
\leq & \frac{8}{\eta^2}\left(Ee^{|\eta| Y^{(t)2}_{i}} + Ee^{|\eta| Y^{(t)2}_{j}}\right)e^{|\eta r_{ijt}|} \leq C_4
\end{split}
\end{equation*}
where $C_4$ is a constant which does not depend on $n_1, n_2, p$. Thus, we set $\bar B_n^2 = n_tC_1$; based on lemma 1 in \cite{Cai_Liu}, we have
\begin{equation}\label{ineq:lemma_1}
\begin{split}
{\rm pr}\left(\left|\frac{1}{n_t}\sum_{k=1}^{n_t}(Y_{ik}^{(t)}Y_{jk}^{(t)} - r_{ijt})\right| \geq C_{\eta/2}C_4^{1/2} x \right) \leq \exp(-n_tx^2). 
\end{split}
\end{equation}
for all $0 < x \leq C_1^{1/2}$, where $C_{\eta/2} = \eta/2+2/\eta$. Next for $\bar Y_i^{(t)}$, we similarly apply Lemma 1 in \cite{Cai_Liu} and get
\begin{equation}\label{ineq:lemma_2}
{\rm pr}\left(|\bar Y^{(t)}_i| \geq C_5x\right) \leq \exp(-n_tx^2)
\end{equation}
for all $0 < x\leq C_5^{1/2}$. Combining \eqref{ineq:lemma_1} and \eqref{ineq:lemma_2},
\begin{equation}\label{ineq:lemma_3}
{\rm pr}\left(\max_{ij} \left|\frac{1}{n_t}\sum_{k=1}^{n_t}(Y_{ik}^{(t)}Y_{jk}^{(t)} - r_{ijt})\right| \leq C_{\eta/2}C_4^{1/2} x\quad \text{and} \quad \max_{i,t}|\bar Y_i^{(t)}| \leq C_5x \right) \leq 1 - 2p^2\exp(-n_tx^2)
\end{equation}
for all $0 < x \leq \min\left(C_1^{1/2}, C_5^{1/2}\right)$. Finally, \eqref{eq:hatsigma_ijt - sigma_ijt} and \eqref{ineq:lemma_3} yield \eqref{ineq:main_condition11}.
\quad $\square$

{\noindent \bf Proof of Theorem~\ref{pr:lower_bound}.} Without loss of generality, we assume $n_1 \leq n_2$. For $(\R_1, \R_2) \in \mathcal{G}_q(s_0(p))$, set $\SSigma_2=\R_2=I_{p\times p}$ and we have already known this information. The estimation of sparse difference immediately becomes the estimation of the sparse correlation matrix $\R_1$. Then the lower bound result for estimating single sparse covariance matrix can be used to prove this theorem.

We follow the idea of \cite{CZ12} and define the set of diagonal-1 covariance matrices as
$$\mathcal{F}_q(s_0(p)) = \left\{\SSigma: \SSigma\succeq 0, \diag(\SSigma) = 1, \max_i\sum_{j=1}^p|\sigma_{ij}|^q \leq s_0(p) \right\}.$$
We have $\left\{(\R_1, \I): \R_1 \in \mathcal{F}_q(s_0(p))\right\} \subseteq \mathcal{G}_q(s_0(p))$. Besides, the proof of Theorem 2 in \cite{CZ12} shows that
\begin{equation}
\inf_{\hat\SSigma} \sup_{\SSigma \in \mathcal{F}_q(s_0(p))} E\|\hat{\SSigma} - \SSigma\|^2 \geq Cs_0(p)\left(\frac{\log p}{n}\right)^{1-q}
\end{equation}
Since the correlation matrix equals to covariance matrix (i.e. $\R = \SSigma$) when $\diag(\SSigma) = 1$, then
\begin{equation}
\begin{split}
& \inf_{\hat\D} \sup_{(\R_1, \R_2)\in \mathcal{G}_q(s_0(p))} E\| \hat{\D} - (\R_1 - \R_2)\|^2\\
 \geq & \inf_{\hat \D} \sup_{(\R_1, \I): \R_1 \in \mathcal{F}_q(s_0(p))} E\|\hat{\D} - (\R_1 - \I)\|^2\\
\geq & \inf_{\hat\R_1} \sup_{\R_1 \in \mathcal{F}_q(s_0(p))} E\|\hat{\R}_1 - \R_1\|^2\\
\geq & \inf_{\hat\SSigma} \sup_{\SSigma_1\in \mathcal{F}_q(s_0(p)),  \diag(\SSigma_1) =1} E\|\hat{\SSigma}_1 - \SSigma_1\|\\
\geq & Cs_0^2(p)\left(\frac{\log p}{n_1}\right)^{1-q} \geq  \frac{C}{2}s_0^2(p)\left(\frac{\log p}{n_1} + \frac{\log p}{n_2}\right)^{1-q}
\end{split}
\end{equation}
which implies \eqref{ineq:lower_bound_spe}. By $\|\cdot\|_{\ell_1} \geq \|\cdot\|$ for symmetric matrices, \eqref{ineq:lower_bound_l1} also follow immediately.

Similarly, \eqref{ineq:lower_bound_F} follows from Theorem 4 of \cite{CZ12}.
\quad $\square$

{\noindent\bf Proof of Proposition~\ref{pr:cor_single}.} The proof of Proposition \ref{pr:cor_single} is similar to Theorem \ref{th:main_cor}. For the upper bound, again, we split the whole events into three,
\begin{equation}
A_1 = \{\max_{ij} \frac{|\hat{\sigma}_{ij} - \sigma_{ij}|}{\sqrtp{\log p \hat{\theta}_{ij}/n}} \leq \tau/4+3, \quad \text{and}\quad \max_{ij} \frac{|\hat{\theta}_{ij} - \theta_{ij}|}{\sigma_{ii}\sigma_{jj}} \leq \varepsilon\},
\end{equation}
\begin{equation}
\begin{split}
A_2 = \Bigg\{& \max_{ij}\frac{|\hat{\sigma}_{ij} - \sigma_{ij} |}{\sqrtp{\log p\hat{\theta}_{ij}/n}} > \tau/4+3, \quad \max_{ij} \frac{|\hat{\theta}_{ij} - \theta_{ij}|}{\sigma_{ii}\sigma_{jj}} \leq \varepsilon\\
& \text{and} \quad \max_{ij} \frac{|\hat{\sigma}_{ij} - \sigma_{ij}|}{\sqrtp{\sigma_{ii}\sigma_{jj}}} \leq \min(0.5, C_1 C_3)\Bigg\}
\end{split}
\end{equation}
\begin{equation}
A_3 = (A_1 \cup A_2)^c.
\end{equation}
Here $\varepsilon$ is the fixed constant which satisfies $0<\varepsilon < \nu_0/2$ where $\nu_0$ was introduced in \eqref{ineq:var_yy}; $C_1, C_3$ are constants specified in Lemma \ref{lm:Cai_Liu_lemma2}. 
Similarly to the proof of Theorem \ref{th:main_cor}, we can prove the following statements.
\begin{enumerate}
\item Under $A_1$, 
$$\|\hat{\R}^\ast-\R\|^2 \leq Cs_0^2(p)\left(\frac{\log p}{n}\right)^{1-q},$$
$$\|\hat{\R}^\ast-\R\|^2_{\ell_1} \leq Cs_0^2(p)\left(\frac{\log p}{n}\right)^{1-q},$$
$$\|\hat{\R}^\ast -\R \|_F^2 \leq Cs_0(p)\left(\frac{\log p}{n}\right)^{1-q/2}.$$

\item For $A_2$, 
$$\int_{A_2} \|\hat{\R}^\ast - \R\|^2dP \leq C(p^{-\tau/4 + 1}\log p)\frac{1}{n}$$
$$\int_{A_2} \|\hat{\R}^\ast - \R\|_{\ell_1}^2dP \leq C(p^{-\tau/4 + 1}\log p)\frac{1}{n}$$
$$\int_{A_2} \|\hat{\R}^\ast - \R\|_F^2dP \leq C(p^{-\tau/4 + 1}\log p)\frac{1}{n}$$

\item For $A_3$, 
$$\int_{A_3}\|\hat{\R}^\ast - \R\|^2dP \leq C\frac{p^{-M+2}}{n}$$
$$\int_{A_3}\|\hat{\R}^\ast - \R\|_{\ell_1}^2dP \leq C\frac{p^{-M+2}}{n}$$
$$\int_{A_3}\|\hat{\R}^\ast - \R\|_F^2dP \leq C\frac{p^{-M+2}}{n}$$
\end{enumerate}
The rest of proof, including the lower bound results, are omitted here as they are essentially the same as Theorem \ref{th:main_cor}.
\quad $\square$

\end{document}